\documentclass{aa}  
\usepackage{afterpage}
\usepackage{amsmath}
\usepackage{color}
\usepackage{graphicx}
\usepackage{natbib}
\usepackage{pdflscape}  
\usepackage{threeparttable} 
\usepackage{txfonts}
\usepackage{upgreek}
\usepackage{url}
\usepackage{float}
\usepackage[rightcaption]{sidecap}
\usepackage{multirow}
\usepackage{multicol}
\usepackage{ulem} 
\usepackage{stfloats}
\usepackage[colorlinks=true,linkcolor=blue,urlcolor=blue,citecolor=blue]{hyperref}
\bibpunct{(}{)}{;}{a}{}{,} 

\newcommand{\kms}{km\,s$^{-1}$}
\newcommand{\micron}{$\upmu$m}

\newcommand{\pcmmm}{$\,\mathrm{cm}^{-3}$}

\newcommand{\avir}{\alpha_\mathrm{vir}}

\begin{document} 

   \title{CHIMPS: Physical properties of molecular clumps \\ across the inner Galaxy\thanks{Tables \ref{tab:FWcat} and \ref{tab:properties} are available in electronic form via anonymous ftp to \url{cdsarc.u-strasbg.fr} (\url{130.79.128.5}) or via \url{http://cdsarc.u-strasbg.fr/viz-bin/qcat?J/A+A/}}}

   \author{A. J. Rigby
          \inst{1,2}
          \and
          T. J. T. Moore\inst{2} \and
          D. J. Eden\inst{2} \and
          J. S. Urquhart\inst{3} \and
          S. E. Ragan\inst{1} \and
          N. Peretto\inst{1} \and
          R. Plume\inst{4} \and \\
          M. A. Thompson\inst{5} \and
          M. J. Currie\inst{6} \and
          G. Park\inst{7}
          }

   \institute{School of Physics \& Astronomy, Cardiff University, Queen's Building, The Parade, Cardiff, CF24 3AA, UK\\
              \email{rigbya@cardiff.ac.uk}
         \and
             Astrophysics Research Institute, Liverpool John Moores University, IC2, Liverpool Science Park, 146 Brownlow Hill, Liverpool L3 5RF, UK
        \and
            Centre for Astrophysics and Planetary Science, The University of Kent, Canterbury, Kent CT2 7NH, UK
        \and 
            Department of Physics and Astronomy, University of Calgary, 2500 University Drive NW, Calgary, AB T2N 1N4, Canada
        \and
            Centre for Astrophysics Research, Science \& Technology Research Institute, University of Hertfordshire, College Lane, Hatfield, Herts AL10 9AB, UK
        \and
            RAL Space, Rutherford Appleton Laboratory, Harwell Oxford, Didcot, Oxfordshire OX11 0QX, UK
        \and
            Korea Astronomy and Space Science Institute, 776 Daedeokdae-ro, Yuseong-gu, Daejeon 34055, Republic of Korea
             }

   \date{Received XXX; accepted YYY}

   \abstract{The latest generation of high-angular-resolution unbiased Galactic plane surveys in molecular-gas tracers are enabling the interiors of molecular clouds to be studied across a range of environments. The CO Heterodyne Inner Milky Way Plane Survey (CHIMPS) simultaneously mapped a sector of the inner Galactic plane, within $27.8^{\circ} \lesssim \ell \lesssim 46.2^{\circ}$ and $|b| \leq 0 \fdg 5$, in $^{13}$CO (3--2) and C$^{18}$O (3--2) at an angular resolution of 15\,arcsec. The combination of the CHIMPS data with $^{12}$CO (3--2) data from the CO High Resolution Survey (COHRS) has enabled us to perform a voxel-by-voxel local-thermodynamic-equilibrium (LTE) analysis, determining the excitation temperature, optical depth, and column density of $^{13}$CO at each $\ell, b, v$ position. Distances to discrete sources identified by {\sc FellWalker} in the $^{13}$CO (3--2) emission maps were determined, allowing the calculation of numerous physical properties of the sources, and we present the first source catalogues in this paper. We find that, in terms of size and density, the CHIMPS sources represent an intermediate population between large-scale molecular clouds identified by CO and dense clumps seen in thermal dust continuum emission, and therefore represent the bulk transition from the diffuse to the dense phase of molecular gas. We do not find any significant systematic variations in the masses, column densities, virial parameters, mean excitation temperature, or the turbulent pressure over the range of Galactocentric distance probed, but we do find a shallow increase in the mean volume density with increasing Galactocentric distance. We find that inter-arm clumps have significantly narrower linewidths, and lower virial parameters and excitation temperatures than clumps located in spiral arms. When considering the most reliable distance-limited subsamples, the largest variations occur on the clump-to-clump scale, echoing similar recent studies that suggest that the star-forming process is largely insensitive to the Galactic-scale environment, at least within the inner disc.}

   \keywords{molecular data -- surveys -- stars:formation -- ISM:molecules -- ISM:structure -- Galaxy: structure}

   \maketitle


\section{Introduction}
Large-scale and unbiased Galactic-plane surveys in wavelengths from the mid-infrared to millimetre regimes, covering both continuum and line emission have, in the last few decades, enabled significant advances in our knowledge of the connection between molecular-cloud physics and star formation. Molecular-line surveys have found that molecular clouds -- the birthplaces of stars and star clusters -- have velocity dispersions an order of magnitude larger than expected from their thermal properties alone \citep[e.g.][]{Larson81}, a characteristic that is generally interpreted as evidence of supersonically turbulent interiors. Many subsequent observations \citep[e.g.][]{Elmegreen00,Rathborne+09} and simulations \citep[e.g.][]{Klessen+05,Bonnell+11,Padoan+16} support this turbulent picture of molecular-cloud interiors, in which both their formation and dissipation is fast. Recent results from the APEX Telescope Large Area Survey of the Galaxy \citep[ATLASGAL;][]{Urquhart+18} find that evolutionary tracers contained within high-mass dense clumps identified at submillimetre wavelengths also support a picture in which the clumps are assembled rapidly, and the embedded star-formation begins almost immediately.

The role that the kpc-scale Galactic environment plays in star-formation and in determining the physical properties of molecular clouds and their constituent substructures is a matter of ongoing research. In the Central Molecular Zone (CMZ), the star-formation rate is known to be an order of magnitude lower than expected from density-threshold-type star-formation prescriptions \citep[e.g.][]{Longmore+13,Barnes+17,Walker+18}, and is believed to be inhibited by the high turbulent energy densities in that environment \citep{Kruijssen+14,Henshaw+16}. In the disc of the Galaxy, there are good reasons to expect systematic radial variations in the star-forming process and various trends with increasing Galactocentric radius have been measured, including: a decreasing metallicity \citep[e.g.][]{Caputo+01,Luck+Lambert11}, a decreasing interstellar radiation field \citep[e.g.][]{Popescu+17}, decreasing molecular-to-atomic gas ratio \citep{Sofue+Nakanishi16}, and increases in the dust temperature \citep{Urquhart+18} and gas-to-dust mass ratio \citep{Giannetti+17}. \citet{Ragan+16} found that the fraction of star-forming \textit{Herschel} sources -- distinguished by the presence of a 70\,\micron\ compact source -- also shows a modest decline with increasing Galactocentric radius.

Several studies using Galactic plane surveys have found no enhancement in various tracers of star-forming activity associated with spiral arms that might support a triggering scenario, such as the clump-formation efficiency \citep[or the `dense-gas mass fraction'][]{Eden+12,Eden+13} or the star-formation efficiency \citep[e.g.][]{Moore+12,Eden+15}, and no evidence for systematic age gradients across them \citep{Ragan+18}. In smoothed particle hydrodynamics simulations, \citet{Duarte-Cabral+Dobbs16} also find little difference between the mean properties of giant molecular clouds (GMCs) in arm and inter-arm regions, though the tails of the some distributions do show differences. These studies suggest that the key link between spiral arms and star-formation rate is through the assembly of large reservoirs of molecular gas and clouds, but without significantly changing the physics of that process compared with inter-arm regions. 

The majority of the mass in molecular clouds consists of molecular hydrogen (H$_2$). The clouds are typically extremely cold, with gas temperatures of $\sim$10--20\,K \citep[e.g.][]{Bergin+Tafalla07,Roman-Duval+10} and, in such environments, H$_2$ is unable to efficiently radiate via its least-energetic electric-quadrupole transitions. The second-most-abundant molecule in the ISM, carbon monoxide (CO), however, emits readily at such temperatures via its lowest rotational dipole transitions and observations of CO emission are, therefore, able to probe molecular-cloud physics. The Galactic Ring Survey \citep[GRS;][]{Jackson+06} -- a survey of $^{13}$CO (1--0) emission covering some 74.5 deg$^{2}$ of the Northern Inner Galactic plane at 46-arcsec resolution -- provided a benchmark in high-resolution, unbiased spectral imaging that has been invaluable over the last decade, providing a view of the molecular counterpart to the ever-increasing volume of thermal-dust-continuum surveys.

In the Northern Sky, the CO Heterodyne Inner Milky Way Plane Survey \citep[CHIMPS;][]{Rigby+16} and the CO High-Resolution Survey \citep[COHRS;][]{Dempsey+13}, both carried out at the 15-m James Clerk Maxwell Telescope (JCMT), have covered much of the GRS survey area and provide a higher-angular resolution (15 arcsec) view of the $J$=3--2 transition of the three most common CO isotopologues: $^{12}$CO, $^{13}$CO, and C$^{18}$O. More recently the FOREST unbiased Galactic-plane Imaging survey with the Nobeyama 45-m telescope \citep[FUGIN;][]{Umemoto+17} has been completed, providing a view of $^{12}$CO, $^{13}$CO, and C$^{18}$O $J$=1--0 emission in the Northern Galactic plane at 20-arcsec resolution. The Southern Galaxy is also becoming increasingly well surveyed in CO, with coverage in the $J$=1--0 transition of $^{12}$CO, $^{13}$CO, and C$^{18}$O from the Three-mm Ultimate Mopra Milky Way Survey \citep[ThrUMMS;][]{Barnes+15} and the Mopra Southern Galactic Plane CO Survey (\citealt{Braiding+15}, also in C$^{17}$O), and the Structure, Excitation, and Dynamics of the Inner Galactic Interstellar Medium survey \citep[SEDIGISM;][]{Schuller+17} covering the $J$=2--1 transition of $^{13}$CO and C$^{18}$O . The coverage of congruent CO survey data in multiple isotopologues and transitions enables more complete and large-scale analyses of the excitation conditions of molecular clouds than has ever been possible.

In this paper, we use a local-thermodynamic-equilibrium (LTE) model to combine data from the CHIMPS and COHRS surveys in order to determine the physical conditions of the interiors of a large sample of molecular clouds at high resolution. In Sect.~\ref{sec:obs}, we describe the observations used followed by a description of our LTE methodology in Sect.~\ref{sec:LTE}. The source extraction and subsequent distance assignments are described in Sect.~\ref{sec:FW+distances}, and we determine the physical properties of these sources in Sect.~\ref{sec:properties}. We discuss our findings in Sect.~\ref{sec:discussion}, and make our concluding remarks in Sect.~\ref{sec:conclusion}.

\section{Data} \label{sec:obs}

CHIMPS \citep{Rigby+16} is a survey of the $J$=3--2 rotational transition of $^{13}$CO and C$^{18}$O, covering approximately 19 square degrees of the Galactic plane in the longitude range $27\fdg 8 \lesssim \ell \lesssim 46 \fdg 2$ and latitudes of $|b| < 0\fdg 5$. The observations were carried out at the 15-m JCMT in Hawaii which, at 330\,GHz, has an angular resolution of 15 arcsec. The Heterodyne Array Receiver Program (HARP) was used in conjunction with the Auto-Correlation Spectral Imaging System (ACSIS) backend \citep{Buckle+09} to observe the two isotopologues simultaneously, with a binned channel width of $0.5$\,\kms\ and a bandwidth of $\sim 200$\,\kms\ in velocity. The band centroid varies with longitude to follow the spiral arms, covering a line-of-sight velocity range of $-50$ to $+150$\,\kms\ at the lowest longitudes, and $-75$ to $+125$\,\kms\ at the higher-longitude end of the survey region. The survey achieved mean rms sensitivities of $\sigma(T_\mathrm{A}^{*}) \approx 0.6\,$K and $0.7$\,K per 0.5-\kms\ velocity channel for $^{13}$CO (3--2) and C$^{18}$O (3--2), respectively, though the sensitivity changes across the survey region due to varying weather conditions, and varying numbers of working HARP receptors. In $^{13}$CO (3--2), the rms of individual cubes ranges between $\sigma(T_\mathrm{A}^{*})=0.37\,$K and $1.51\,$K per channel, and between $\sigma(T_\mathrm{A}^{*})=0.43\,$K and 1.77\,K per channel in C$^{18}$O (3--2).

COHRS \citep{Dempsey+13}, also carried out at the JCMT, is a spectral survey of $^{12}$CO (3--2) emission within the first Galactic quadrant. The first data release covered a longitude range of $17\fdg 5 \leq \ell \leq 50 \fdg 25$ and with a latitude coverage of  $|b| \leq 0 \fdg 25$, and two small segments with $|b| \leq 0\fdg5$. In this paper, we also make use of further COHRS data that have been observed in the intervening time period, expanding the latitude coverage to the full $|b| \leq 0\fdg5$ throughout the CHIMPS survey range, and will be presented in an upcoming paper (Park et al. in prep.). The COHRS data have an effective angular resolution of $16.6$ arcsec and, with 1\,\kms\ spectral bins, reach an rms sensitivity of $\sigma(T_\mathrm{A}^{*}) \approx 1$\,K per channel. 

For the analysis presented in the following section, the COHRS data were re-gridded using the Starlink software package \citep{Currie+14} -- specifically the {\sc kappa} routine {\sc wcsalign} -- to match the CHIMPS voxel grid of of $7.6 \mathrm{\,arcsec} \times 7.6 \mathrm{\,arcsec} \times 0.5$\,\kms, using a linear interpolation to upsample the spectral data. The data from both surveys were also converted from the corrected antenna-temperature scale, $T_\mathrm{A}^{*}$, to the main-beam brightness temperature scale, using $T_\mathrm{mb} = T_\mathrm{A}^{*} / \eta_\mathrm{mb}$, adopting main-beam efficiencies of $\eta_\mathrm{mb} =0.72$ and 0.61 for the CHIMPS and COHRS data, respectively \citep{Buckle+09}. The data have also been spatially smoothed to a common resolution of 27.4 arcsec (resulting from the application of a 3-pixel FWHM Gaussian smooth to the CHIMPS data) in order to increase the signal-to-noise ratio (S/N). The smoothed data have rms values of $0.12^{+0.14}_{-0.03}$ K for the $^{12}$CO (3--2) data,  $0.09^{+0.03}_{-0.03}$ K for the $^{13}$CO (3--2) data, and $0.13^{+0.02}_{-0.02}$ for the C$^{18}$O (3--2) data per 0.5\,\kms\ channel, where the quoted values correspond to the median of the distribution, with uncertainties quoted as the first and third quartiles.

\vspace{-2mm}
\section{Local Thermodynamic Equilibrium analysis} \label{sec:LTE}

The combination of CO survey data allows us to calculate the excitation temperature, optical depth, and column density of each voxel within the CHIMPS data. In the following analysis, an assumption that the molecular gas can be described as a system in LTE is adopted. The brightness temperature of an isothermal slab of CO radiating at a frequency $\nu$ is given by:

\vspace{-1mm}
\begin{equation} \label{eq:Tb}
    T_\mathrm{B}(\nu) = J(\nu) \, (1 - e^{-\tau_\nu}).
\end{equation}

\noindent Here we assume that the brightness temperature of the emitting gas can be measured by the main-beam brightness temperature $T_\mathrm{mb}$, and

\begin{equation} \label{eq:J}
    J(\nu) = \frac{h \nu}{k_\mathrm{B}} \left(\frac{1}{e^{h \nu/k_\mathrm{B} T_\mathrm{ex}} - 1} - \frac{1}{e^{h \nu / k_\mathrm{B} T_\mathrm{bg}}-1} \right),
\end{equation}

\noindent where $T_\mathrm{ex}$ is the excitation temperature of the line, $T_\mathrm{bg}$ is the temperature of the cosmic microwave background, which has a value of 2.7\,K \citep{Fixsen09}, $\tau_\nu$ is the optical depth and $k_\mathrm{B}$ is Boltzmann's constant. 

In the following, we largely adopt a standard approach as outlined in \citet{WilsonRohlfs+Huttemeister}, which reasonably assumes that all $^{12}$CO (3--2) emission is optically thick in order to determine the excitation temperature ($T_\mathrm{ex}$). Subsequently, the optical depth ($\tau_\nu$) of the less opaque $^{13}$CO (3--2) emission is calculated, enabling the determination of the $^{13}$CO column density. We apply this methodology to determine these quantities in a similar manner to \citet{Roman-Duval+10}, where each quantity is determined for each $\ell, b, v$ position or `voxel' (i.e. three-dimensional pixel) within the survey volume. This has the advantage that any following derived source properties are independent of the segmentation method used to extract the sources, compared with performing the analysis to velocity-integrated properties. One drawback of performing this analysis on a voxel-by-voxel basis is that any self-absorption in $^{12}$CO or $^{13}$CO (3--2) is largely unaccounted for, and while we perform a first-order adjustment of our method with respect to the $^{12}$CO (3--2) in Sect.~\ref{sec:Tex}, we do not see evidence for significant self-absorption in $^{13}$CO (3--2) throughout the CHIMPS survey. 

We adopt a short-hand notation in which $T_{12}$, $T_{13}$ and $T_{18}$ refer to the main-beam brightness temperatures of $^{12}$CO (3--2), $^{13}$CO (3--2), and C$^{18}$O (3--2), respectively. We also define a similar short-hand notation for the the abundance ratios of $^{12}$CO, $^{13}$CO, and C$^{18}$O with respect to H$_2$, which we define as $R_{12}$, $R_{13}$ and $R_{18}$, respectively. Since the full CHIMPS survey is too large to mosaic into a single data cube for our analysis, the survey was mosaicked into ten large cubes that we call `Regions', which are described in further detail in Sect.~\ref{sec:sourceextraction}. The following calculations were, therefore, performed on our ten Region mosaics, for which we have a data cube for each of the isotopologues.

\subsection{Excitation temperature}\label{sec:Tex}

By assuming that the optical depth of $^{12}$CO (3--2) is much greater than unity wherever it is detected, the excitation temperature can be calculated from Eq.~\ref{eq:Tb}:

\begin{equation}
    T_\mathrm{ex} = 16.6 \left[ \ln\left(1 + \frac{16.6}{T_{12} + 0.04}\right) \right]^{-1}.
\end{equation}

In practice, this does not provide a good solution to the excitation temperature in all positions, because there are places in which $T_{12} < T_{13}$, which may be a result of self-absorption, or strong gradients in density or gas temperature that are better traced by $^{13}$CO than $^{12}$CO. If there is a robust detection of C$^{18}$O (with $\mathrm{S/N} > 5$) in these regions, then we adopt a similar excitation-temperature formulation, but derived from an assumption that the $^{13}$CO (3--2) is optically thick, or else the excitation temperature is undefined.

\subsection{Optical depth}

Once the excitation temperature in a given voxel has been defined, the optical depth is also calculated from Equation~\ref{eq:Tb}:

\vspace{-2mm}
\begin{equation} \label{eq:opticaldepth}
    \tau_{13} = -\ln \left[ 1 - \frac{T_{13}}{15.9} \left( \frac{1}{e^{15.9 / T_\mathrm{ex}}-1} - 0.0028 \right)^{-1} \right].
\end{equation}
    
\noindent In regions where the $T_{12} < T_{13}$, the excitation temperature was calculated from $T_{13}$ by assuming that $^{13}$CO emission is optically thick, and we adapt Eq.~\ref{eq:opticaldepth} to calculate the optical depth of C$^{18}$O (3--2). The optical depth of C$^{18}$O was then used to estimate $\tau_{13} = \tau_{18} R_{13}/R_{18}$ by adopting an abundance ratio of $R_{13}/R_{18} = 6.5$ \citep{Wilson+Rood94} at a Galactocentric distance of 5.5\,kpc which represents the median distance within our sample.

\subsection{Column density} \label{sec:columndensity}

Once the excitation temperature and optical depth of each $^{13}$CO (3--2) voxel has been determined, the total column density can be determined by calculating the column density within a specific $J$ energy level, and multiplying by a partition function, $Z$, which is the sum over all states:

\vspace{-2mm}
\begin{equation} \label{eq:Ntotal}
    N_{13}(\mathrm{total}) = N_{13}(J)\frac{Z}{2J + 1} \exp{\left[\frac{h B J (J+1)}{k_\mathrm{B} T_\mathrm{ex}}\right]}.
\end{equation}

\noindent Assuming that vibrationally excited states are not populated, the partition function may be approximated as:

\vspace{-2mm}
\begin{equation} \label{eq:Z}
    Z \approx \frac{k_\mathrm{B}}{hB} \left( T_\mathrm{ex} + \frac{hB}{3k_\mathrm{B}} \right).
\end{equation}

\noindent The column density within the $J=2$ state, in units of cm$^{-2}$, is calculated as:

\vspace{-2mm}
\begin{equation} \label{eq:Nlower}
    N_{13}(J=2) = \frac{8 \pi}{c^3} \frac{g_2}{g_3} \frac{\nu^3}{A_{32}}\frac{1}{1-\exp{(-h\nu / k_\mathrm{B} T_\mathrm{ex})}} \int \tau_\nu \mathrm{d} \varv ,
\end{equation}

\noindent where $g_2$ and $g_3$ are the statistical weights of the $J=2$ and $J=3$ rotational energy levels, $A_{32} = 2.181 \times 10^{-6}$\,s$^{-1}$ is the Einstein A coefficient for the $^{13}$CO (3--2) transition (Leiden Atomic and Molecular Database\footnote{\url{https://home.strw.leidenuniv.nl/~moldata/CO.html}}, \citealt{Schoier+05}), and the integral is over the linewidth with velocity element $\mathrm{d} \varv$. The rotation constant is calculated as $B = h / (8 \pi^2 \mathcal{I})$ where the moment of inertia, $\mathcal{I} = \mu R_\mathrm{CO}^2$, is equal to the reduced mass, $\mu$, multiplied by the mean atomic separation of $R_\mathrm{CO} = 0.112$\,nm. 

\begin{figure*}
  \centering
  \includegraphics[width=\linewidth]{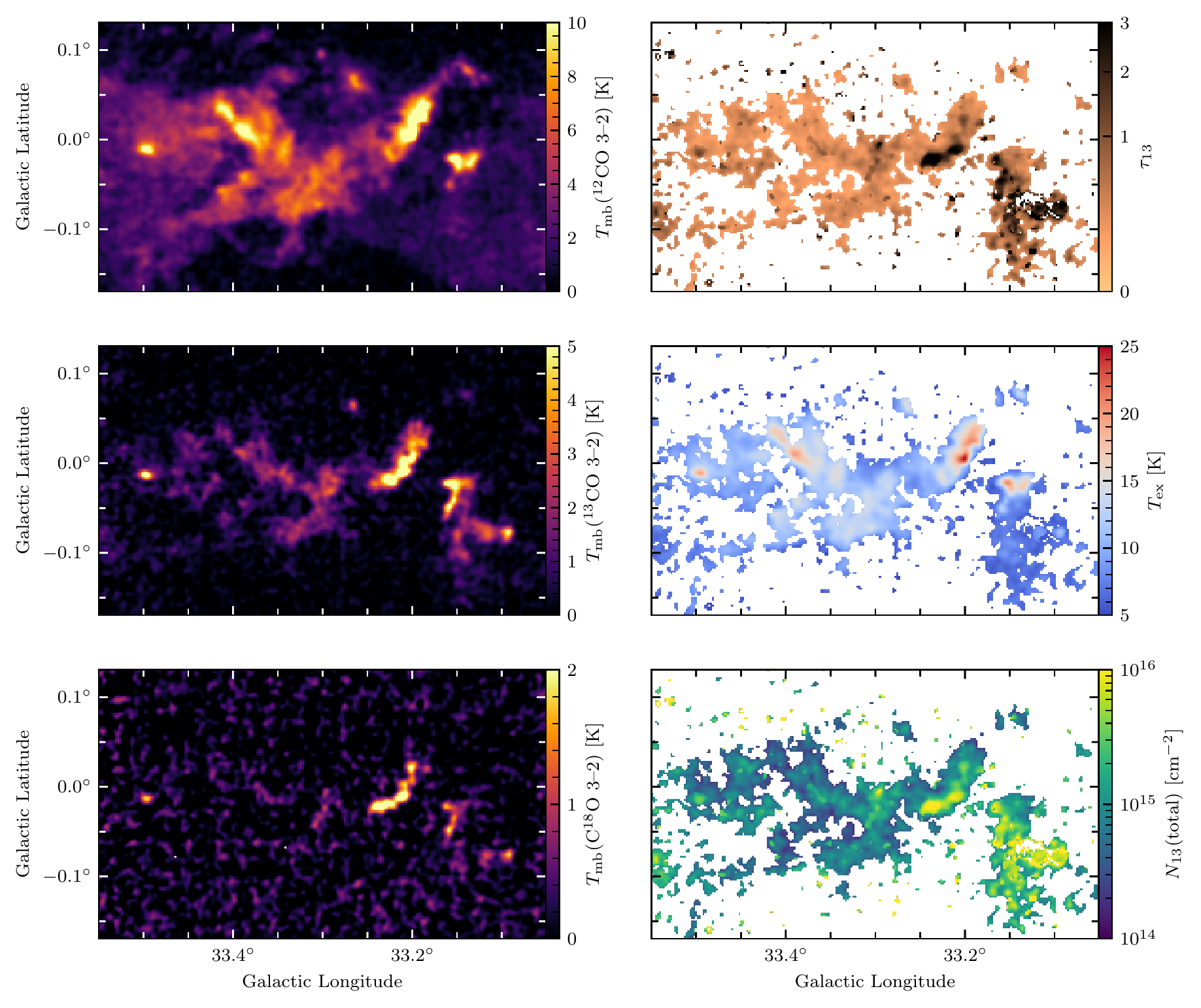}
  \vspace{-5mm}
  \caption{A demonstration of the LTE methodology used in this study for a region centred on $\ell$=33.3\degr, $b$=-0.02\degr\ in the 101.22\,km\,s$^{-1}$ velocity channel. In the left column, the intensities of the three CO emission lines: COHRS $^{12}$CO (3--2) (top panel), CHIMPS $^{13}$CO (3--2) (middle panel), and CHIMPS C$^{18}$O (3--2) (bottom  panel) are shown. In the right column, the derived LTE properties: the $^{13}$CO (3--2) optical depth (top panel), the excitation temperature (middle panel), and the total $^{13}$CO column density (bottom panel) are also shown.}
  \label{fig:LTEexample}
\end{figure*}

We note that the small difference between the values of the partition function used here, and that presented on the Cologne Database for Molecular Spectroscopy\footnote{\url{https://cdms.astro.uni-koeln.de/cdms/portal/}} \citep{Endres+16} is due the hyperfine splitting of $^{13}$C, which we do not account for. The difference in the resultant column densities is 0.5--2\% over a temperature range of 5--20\,K, with the largest discrepancies at the lowest temperatures, and so its impact is considered to be negligible for our purposes.

In Fig.~\ref{fig:LTEexample} we demonstrate this method by showing a single velocity slice of a region in the three CO emission lines, alongside the derived excitation temperature, optical depth and total $^{13}$CO column density slices.

\vspace{-1.5mm}
\subsection{Uncertainties on LTE properties} \label{sec:uncertainties}
\vspace{-0.5mm}

A bootstrapping approach was used to estimate the uncertainties on the $T_\mathrm{ex}$, $\tau_{13}$ and $N_{13}$(total) voxel values that might arise as a result of the calibration uncertainty on the HARP instrument. The mean difference between typical intensities measured with HARP, and the reported standard values of JCMT calibrators is around $15\%$ \citep{Buckle+09}, and so we multiply the intensity values of each input cube by a factor that is drawn from a normal distribution with a standard deviation mean of 1.00 and a standard deviation of 0.15. The input $^{12}$CO (3--2) values are multiplied by a different randomly generated factor than is used for the $^{13}$CO (3--2) and C$^{18}$O (3--2), while the same factor was adopted for the latter two data cubes since they are always observed simultaneously. We performed 50 realisations of the modified data, and finding that the range of output values per voxel had standard deviations of approximately 5\%, 23\% and 28\% for the excitation temperatures, optical depths, and total column densities, respectively.

\vspace{-1.5mm}
\subsection{Sub-thermal emission}
\vspace{-0.5mm}

\begin{figure*}
  \centering
  \includegraphics[width=\linewidth]{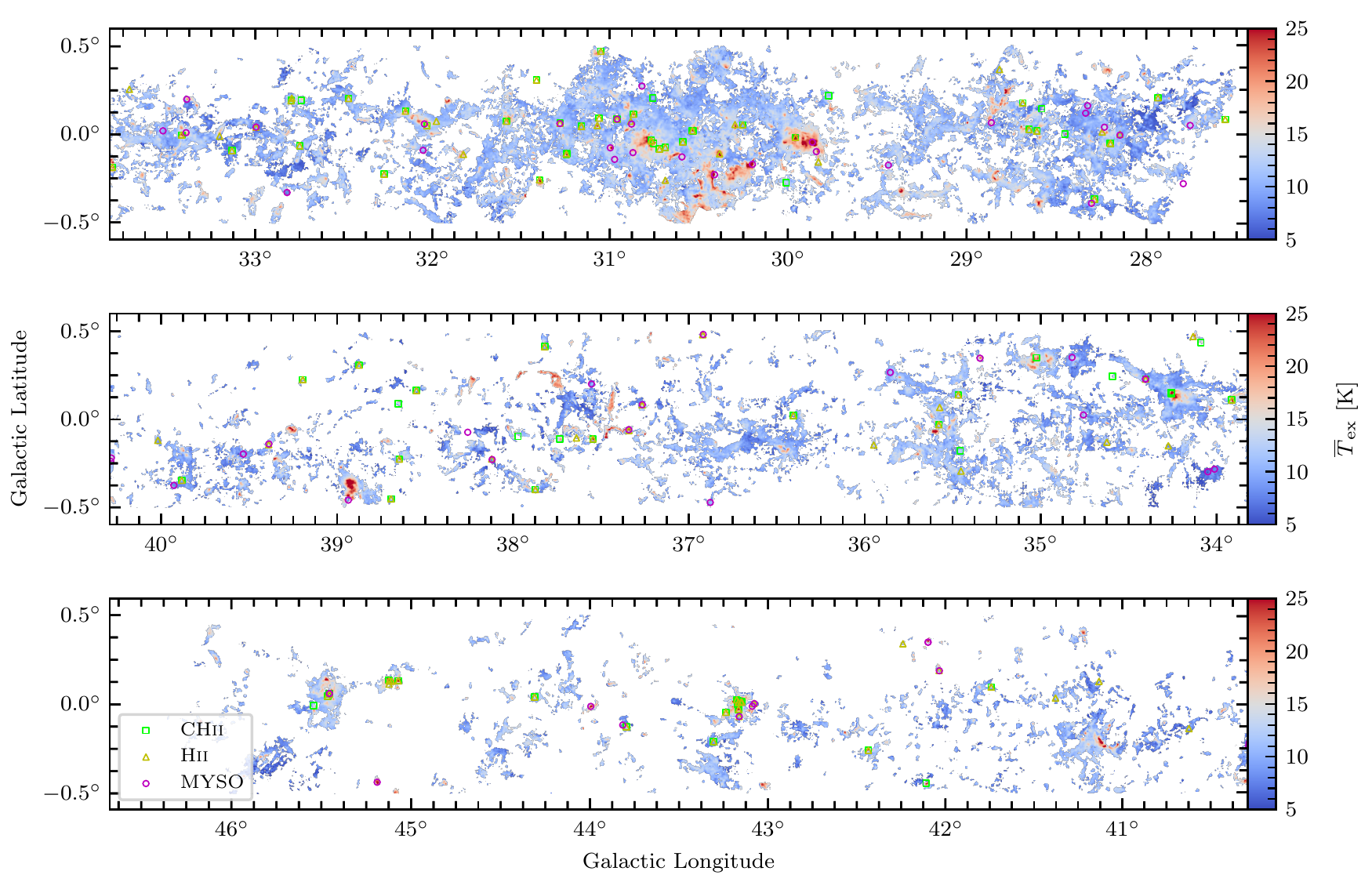}
  \caption{Weighted mean excitation temperature map for CHIMPS. For each pixel, the weighted mean excitation temperature, $\overline{T}_\mathrm{ex}$, has been calculated for the corresponding spectrum, with weightings given by the $T_\mathrm{A}^{*}$ of $^{13}$CO (3--2), and excluding all voxels for which $\mathrm{S/N} <5$. Compact \ion{H}{II} regions from CORNISH \citep{Kalcheva+18} are overlaid as green squares, and \ion{H}{II} regions and MYSOs from the RMS survey \citep{Lumsden+13} are shown as yellow triangles and magenta circles, respectively.}
  \label{fig:Tex_map}
 \end{figure*}

The analysis presented in this section assumes that LTE applies in all voxels in which $^{13}$CO (3--2) is detected. However, gas lying at densities below the critical density of $^{13}$CO (3--2) ($\approx 10^4$\,\pcmmm) will be warmer than the estimated excitation temperature, but may still emit in a sub-thermal mode in which the energy level populations do not follow the Boltzmann distribution. This underestimate in the gas temperature will lead to overestimates in the column density, according to Eq.~\ref{eq:Ntotal}. In Sect.~\ref{sec:properties}, the distribution of mean excitation temperatures of the molecular clumps extracted from the survey is found to have a mean value of 11.5\,K, which matches the expectation for molecular structures covering the size regime from cores, through clumps, to clouds \citep[e.g.][]{Bergin+Tafalla07}. Sub-thermal emission can therefore be assumed not to be a dominant effect here. The effects of sub-thermal emission upon the reported properties will be studied in a future paper.

\subsection{Temperature and column-density maps}

\begin{figure*}
  \centering
  \includegraphics[width=\linewidth]{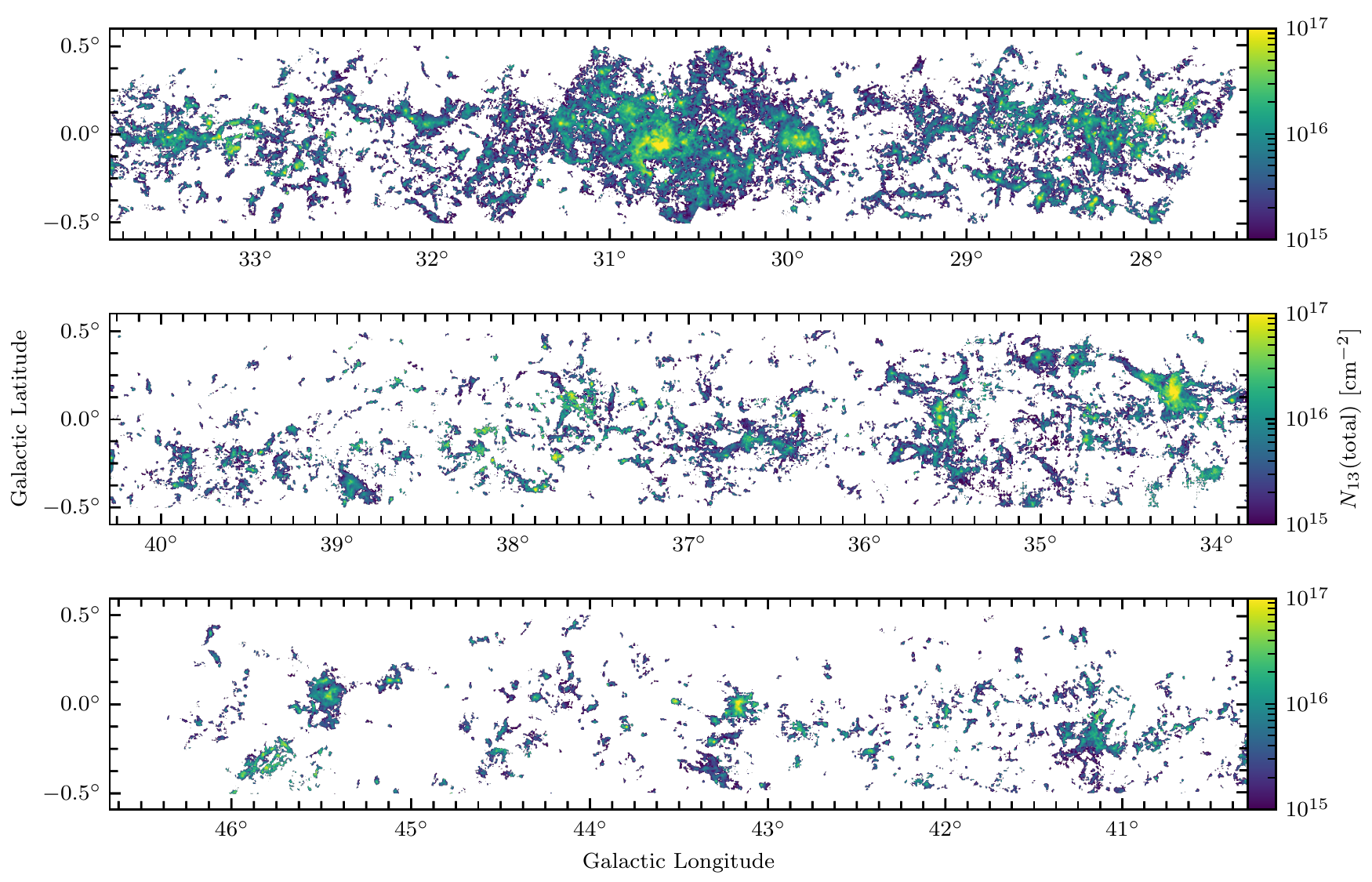}
  \vspace{-5mm}
  \caption{Total $^{13}$CO column density integrated at each position, displayed with a logarithmic intensity scale. The H$_2$ column density is roughly a factor of $10^{6}$ greater.}
  \label{fig:Nco_map}
\end{figure*}

These calculations have enabled one of the first maps of the excitation temperature of molecular gas for a significant region of the Galactic plane to be produced. A map of the excitation temperature across the survey area can be found in Fig.~\ref{fig:Tex_map}, in which each pixel is the mean excitation temperature ($\overline{T}_\mathrm{ex}$), weighted by the intensity of $^{13}$CO (3--2) emission in the spectrum at that position. The mean excitation temperature taken across all pixels is 11.2 K, and the median temperature is $10.8^{+1.7}_{-1.5}$ K, where the uncertainties give the range covering the first and third quartiles, though the temperature becomes as high as $\sim 50$ K towards the centres of intense star formation such as  W43($\ell=30.8$\degr, $b=0.0$\degr), W49 ($\ell= 43.2$\degr, $b=0.0$\degr), and the G34 region ($\ell=34.3$\degr, $b=0.1$\degr). A temperature gradient is visible across the filament located at $\ell=37.4$\degr, $b=-0.1$\degr, which was highlighted in \citet{Rigby+16}. This temperature gradient would appear to add weight to the idea that this filament is an expanding bubble rim, since it is warmer on the inside edge.

The positions of compact \ion{H}{II} regions (C\ion{H}{II}) from CORNISH \citep{Kalcheva+18}, and \ion{H}{II} regions and massive young stellar objects (MYSOs) from the RMS survey \citep{Lumsden+13} are also shown on this figure. About 75\% of the \ion{H}{II} and C\ion{H}{II} regions within the survey area are coincident with an obvious enhancement in $\overline{T}_\mathrm{ex}$, while only 60\% of the MYSOs are, and some of the latter are also quite weak rises in $\overline{T}_\mathrm{ex}$. This makes sense if the heating from the MYSOs is likely to be relatively low-level, although we note that along sightlines with multiple components of $^{13}$CO (3--2), the apparent $\overline{T}_\mathrm{ex}$ could be artificially suppressed, and any optically thin regions of $^{12}$CO (3--2) may lead to further suppression. A few of the unassociated C\ion{H}{II} regions seem to be displaced from the CO altogether, so may be extragalactic or planetary nebulae, or are associated with CO for which we lack sufficient sensitivity.

Figure~\ref{fig:Nco_map} shows a map of the total $^{13}$CO column density, summed over the velocity axis, on a logarithmic intensity scale. The morphology of this map is largely similar to the map of $^{13}$CO (3--2) emission in \citet{Rigby+16}, with the exception that column density is enhanced by the optical depth in regions of C$^{18}$O (3--2) emission. The column density is illustrated in

\noindent terms of the total $^{13}$CO gas column in order to limit the uncertainty in the various conversion factors required to display the H$_2$ column density. A factor of $10^6$ approximately converts $^{13}$CO to H$_2$ column density, since $R_{12}/R_{13} \sim 100$ \citep[e.g.][]{Wilson+Rood94} and $R_{12} \sim 10^{-4}$ \citep[e.g.][]{Frerking+82}.

\section{Source extraction and distance assignments} \label{sec:FW+distances}
\subsection{Source extraction} \label{sec:sourceextraction}

Discrete sources of coherent three-dimensional ($\ell, b, v$) emission within the $^{13}$CO (3--2) data were extracted using the {\sc FellWalker} algorithm \citep{Berry15}, as described in \citet{Rigby+16}. Since the survey does not have a completely uniform sensitivity \citep[see Fig.~3.][]{Rigby+16}, the source extraction was carried out on S/N cubes as opposed to brightness-temperature cubes, in a similar manner to that employed on continuum data in the JCMT Plane Survey \citep[JPS:][]{Moore+15,Eden+17}. We note that several of the {\sc FellWalker} parameters are defined by their relationship to the background rms which is, by definition, equal to unity for S/N cubes. 

First, the 178 individual $^{13}$CO (3--2) cubes of the full survey were mosaicked into the ten large Region cubes (see Sect.~\ref{sec:LTE}), using the {\sc kappa} task {\sc wcsmosaic}, allowing for a small overlap between adjacent Regions, in order to allow the source extraction to be carried out as consistently as possible with the available computing resources. The Region cubes cover between 0.6 and 3.1 square degrees, and are made up of between 7 and 28 cubes from individual observations, depending upon the local tiling strategy, which varied over the observing campaigns. The overlap between adjacent Regions is always at least one tile-width (20 arcmin) in longitudinal extent, which is significantly greater than the largest source sizes, and so does not result in any artificial source-splitting. The Region cubes were then smoothed to an effective resolution of 27.4 arcsec, increasing the S/N by decreasing the rms noise level as described in Sect.~\ref{sec:obs}.

The initial run of {\sc FellWalker} identified a total of 4999 $^{13}$CO (3--2) sources across the survey, although a number of these sources could be false positives. To mitigate the effects of false positives, a cutout of each CHIMPS clump was visually inspected by three independent reviewers and assigned a reliability flag, for which we list the mean value, rounded to the nearest integer, in the catalogue. A value of 1 was assigned to 526 clumps which appear to be false positives, while a value of 2 was assigned to 805 clumps judged to be dubious in some way, and a value of 3 was assigned to the remaining 3664 clumps that we consider to be robust.

\begin{table*}
\centering
\caption{The CHIMPS $^{13}$CO (3--2) {\sc FellWalker} source catalogue.}
\resizebox{\textwidth}{!}{%
\begingroup
\renewcommand{\arraystretch}{1.2} 
\begin{tabular}{ccccccccccccccc}
\hline \hline
\noalign{\smallskip}
Designation &  Region & ID & $\ell_\mathrm{cen}$ & $b_\mathrm{cen}$ & $v_\mathrm{cen}$ & $\sigma_\ell$ & $\sigma_b$ & $\sigma_v$ & $\int T_\mathrm{A}^{*} dv$ & Peak $T_\mathrm{A}^{*}$ & S/N & $n_\mathrm{pix}$ & $n_\mathrm{vox}$ & Flag  \\
 & & & [$\deg$] & [$\deg$] & [\kms] & [arcsec] & [arcsec] & [\kms] & [K\,\kms] & [K] &  \\
(1) & (2) & (3) & (4) & (5) & (6) & (7) & (8) & (9) & (10) & (11) & (12) & (13) & (14) & (15) \\
\noalign{\smallskip}
\hline
\noalign{\smallskip}
G043.167+00.017 & 7 & 1 & 43.16690 & 0.01717 & 8.69 & 84.1 & 45.2 & 4.82 & 38254.3 & 26.0 & 189.2 & 1361 & 24845 & 3 \\
G034.238+00.115 & 3 & 2 & 34.23761 & 0.11535 & 56.79 & 62.8 & 81.7 & 2.15 & 28569.8 & 25.1 & 139.6 & 1504 & 16721 & 3 \\
G034.248+00.166 & 3 & 1 & 34.24799 & 0.16593 & 57.39 & 113.8 & 66.5 & 2.21 & 25146.0 & 31.9 & 155.0 & 2045 & 20476 & 3 \\
G043.163-00.031 & 7 & 3 & 43.16255 & -0.03076 & 11.44 & 60.0 & 55.1 & 3.87 & 17779.4 & 16.6 & 100.2 & 1055 & 13829 & 3 \\
G029.910-00.059 & 1 & 16 & 29.91009 & -0.05885 & 99.98 & 45.2 & 77.0 & 2.86 & 13873.4 & 19.1 & 76.2 & 1012 & 10587 & 3 \\
G030.838-00.059 & 2 & 17 & 30.83761 & -0.05920 & 96.32 & 63.6 & 61.0 & 3.37 & 12603.3 & 9.6 & 74.5 & 1178 & 13050 & 3 \\
G030.722-00.098 & 2 & 11 & 30.72199 & -0.09832 & 93.79 & 51.8 & 82.3 & 3.59 & 12537.4 & 8.0 & 82.6 & 1321 & 17286 & 3 \\
G029.961-00.015 & 1 & 6 & 29.96112 & -0.01455 & 96.87 & 56.6 & 52.8 & 2.52 & 12291.3 & 21.9 & 93.1 & 833 & 8771 & 3 \\
G030.437-00.235 & 1 & 3 & 30.43714 & -0.23493 & 103.55 & 93.4 & 58.4 & 1.95 & 12169.6 & 12.9 & 137.1 & 1649 & 15029 & 3 \\
G029.860-00.050 & 1 & 9 & 29.85996 & -0.04999 & 99.80 & 50.2 & 59.7 & 2.40 & 11636.9 & 21.4 & 87.8 & 895 & 9374 & 3 \\
\noalign{\smallskip}
\hline
\noalign{\medskip}
\multicolumn{15}{p{1.25\textwidth}}{\textbf{Notes.} The columns detail the following: (1) Source designation; (2) mosaic Region number; (3) pixel value identifying the source in corresponding to the {\sc FellWalker} Region mask; (4--6) centroid coordinates; (7--9) intensity-weighted rms sizes in each dimension; (10) total intensity summed over all voxels in the source; (11) intensity of the brightest voxel; (12) the peak signal-to-noise ratio; (13) number of pixels in the projected $\ell$--$b$ silhouette; (14) total number of $\ell, b, v$ voxels in the source; and (15) reliability flag, as described in the text. Only a portion of the full table is shown here to illustrate its form and content. The full table can be downloaded in a machine-readable format from the Canadian Advanced Network Astronomical Research (CANFAR) archive listed in Appendix \ref{app:dataproducts}.}

\end{tabular}
\endgroup}
\label{tab:FWcat}
\end{table*}

Diffuse sources lying close to the detection threshold can be broken up in a sporadic fashion, and may have highly irregular shapes made up of clusters of disconnected pixels, and sources of this type make up many of the sample of clumps flagged as bad. This category also includes single coherent sources at low S/N which are hard to discern by eye. Sources flagged as `dubious' also often consist of diffuse sources at low S/N that may contain multiple intensity peaks, or an irregular profile. This category also contains what are clearly areas of emission, but that have been segmented in a strange way due to lying on a boundary between tiles with different noise levels. This is an undesirable consequence of carrying out the source extraction on S/N maps, but since these clumps are generally small, only a small fraction of the total sample is affected. These flags broadly correspond to regions of the peak S/N distribution, with 95\% of the false positives occurring with peak $\mathrm{S/N} < 10$, while 95\% of the `dubious' clumps have peak $\mathrm{S/N} < 13$. We make the cutouts of all 4999 clumps available online in PDF format (described in Appendix~\ref{app:dataproducts}).

 We list the extracted $^{13}$CO (3--2) emission properties of the ten sources with the greatest integrated intensities in Table~\ref{tab:FWcat}, and include the full {\sc FellWalker} catalogue in the supporting information. We also include the ten $^{13}$CO (3--2) mosaicked Regions at the native resolution, along with their {\sc FellWalker} assignment masks in the supporting information. We discuss the catalogue completeness for various source sizes in Appendix~\ref{app:completeness}, but note that we determine no single comprehensive completeness limit.

\begin{figure}
  \centering
  \resizebox{\hsize}{!}{\includegraphics{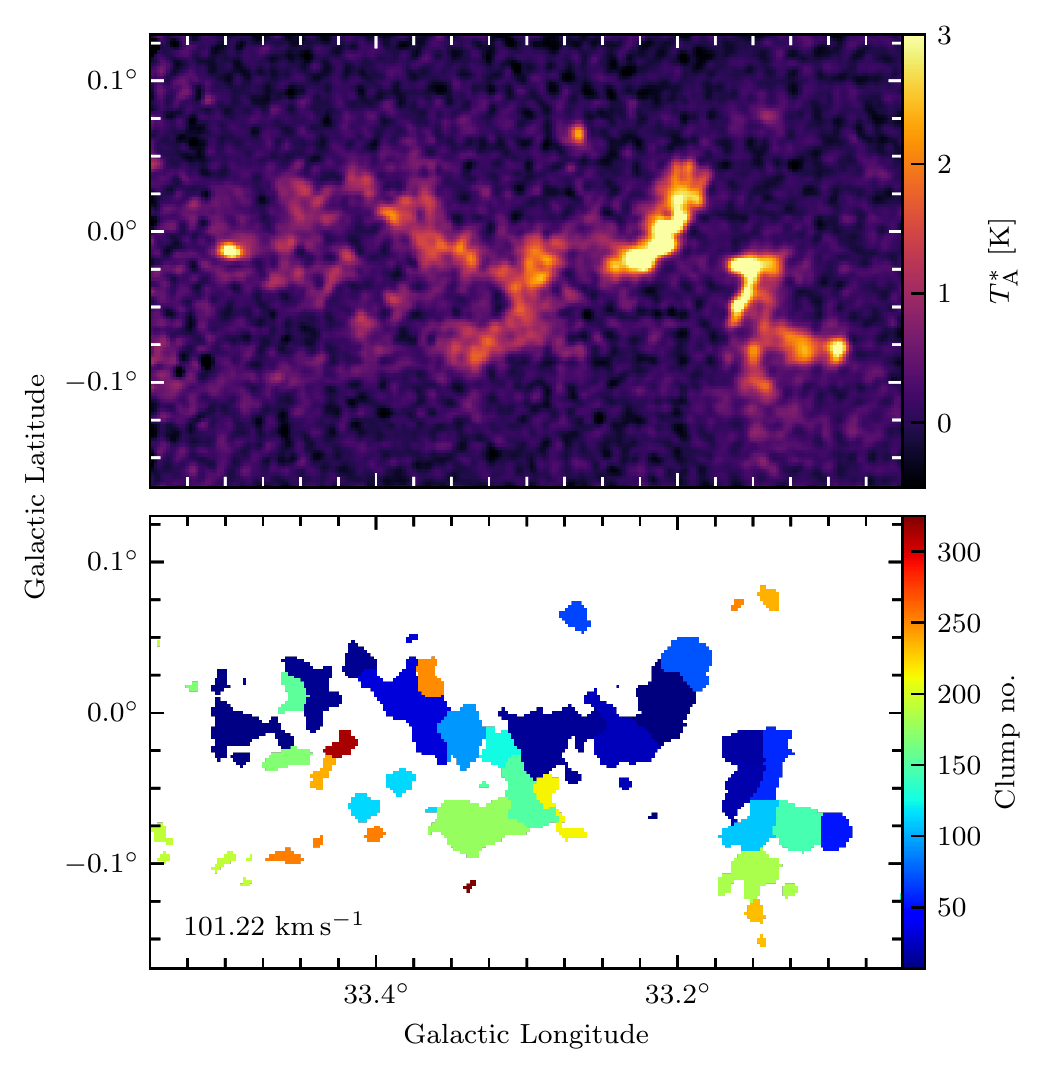}}
  \vspace{-3mm}
  \caption{An example of the {\sc FellWalker} source extraction, taken from Region 3 (see text). The top panel shows $^{13}$CO (3--2) emission in the 101.22\,\kms\ velocity plane at 27.4-arcsec resolution, while the bottom panel shows the corresponding {\sc FellWalker} clump-assignment mask, in which different colours represent different clumps.}
  \label{fig:fellwalker}
\end{figure}

The catalogue contains a Region and ID number for each source, allowing them to be located within the emission cubes using the {\sc FellWalker} masks. Figure~\ref{fig:fellwalker} shows an example of the {\sc FellWalker} clump assignment-mask taken from Region~3, in which each different colour denotes the pixels belonging to a separate catalogued clump, along with the corresponding $^{13}$CO (3--2) emission slice. Some small features exist in this map that do not obviously correspond to real features in the emission map, but which do belong to emission features seen in adjacent channels. This is a result of the {\sc FellWalker} parameters, which allow voxels with $\mathrm{S/N} = 2$ to be included in a clump, so long as they are directly connected to a clump with a peak $\mathrm{S/N} > 5$.

\subsection{Distance assignments} \label{sec:distances}

Establishing the distances to molecular clouds and clumps in the plane of the Milky Way is a fraught process due to the inherent complexity of molecular emission along lines of sight in the Galactic plane. Parallax measurements of distant star-forming regions, such as those acquired through very-long-baseline interferometry \citep[e.g.][]{Reid+14}, represent the most-accurate distance measurements, since they are model-free. However, such measurements do not currently exist in sufficient numbers to apply directly to large survey data such as those of CHIMPS. Establishing kinematic distances to sources with line-of-sight velocity information provides a reasonably robust method. However, such measurements require the assumption of a Galactic rotation curve \citep[e.g.][]{Brand+Blitz93,Reid+14} and an important assumption that objects are obeying purely circular orbits around the Galactic centre.

The Bayesian distance calculator of \citet{Reid+16} was used to estimate the possible near and far kinematic distances -- and associated uncertainties -- for each of the 4999 CHIMPS clumps identified in $^{13}$CO (3--2) emission. The prior that the sources should be associated with spiral arms was removed from the calculation so that the distances are independent of the spiral-arm model. This method adopts the \citet{Reid+14} Galactic rotation model, with a distance to the Galactic centre of $R_0 = 8.34 \pm 0.16$ kpc. 

A variety of methods was used to discriminate between the near and far kinematic distances. In the following, Roman numerals denote the methods used to assign distances which are referred to later in the text. First, a number of distance solutions can be determined through geometric arguments:

\newcounter{List}
\begin{enumerate}[i)]
	\item sources with negative velocities at the positions of their emission peaks were assigned the `far' distance solution because such sources must lie outside the circular orbit described by the Sun about the Galactic centre; 
	\item sources with velocities that exceed the terminal velocity of the Galactic rotation curve at their $\ell, b$ coordinates are assigned the tangential solution, where the near and far distances are equal. The terminal velocity represents the maximum velocity found through purely circular orbits about the Galactic centre, though peculiar velocities of objects close to the tangent points could result in velocities above this value.
    \setcounter{List}{\value{enumi}}
\end{enumerate}

\noindent Next, a series of volumetric searches centred on the $\ell, b, v$ coordinates of the catalogued $^{13}$CO (3--2) clump emission peaks was conducted in order to identify the clumps that are consistent with a distance determination in the literature. Firstly, a search radius of five resolution elements -- 76 arcsec $\times$ 76 arcsec $\times$ 2.5\,\kms\ -- was used to find sources coincident in $\ell, b, v$ with:

\begin{enumerate}[i)]
	\setcounter{enumi}{\value{List}}
	\item ATLASGAL clumps from \citet{Urquhart+18};
	\item ATLASGAL clumps from \citet{Wienen+15};
	\item RMS MYSOs from \citet{Lumsden+13};
	\item the Bolocam Galactic Plane Survey (BGPS) catalogue of \citet{Ellsworth-Bowers+13}; 
	\item BGPS sources with distances determined by \citet{Eden+12} and \citet{Eden+13}; and
	\item GRS clumps identified by \citet{Roman-Duval+09} that are associated with a parent GRS molecular cloud with a known distance. 
    \setcounter{List}{\value{enumi}}
\end{enumerate}    
   
\noindent At this stage, if the designation was unsuccessful, the search volume is expanded, based on new constraints:    
    
\begin{enumerate}[i)]
	\setcounter{enumi}{\value{List}}
	\item A further search for coincident GRS molecular clouds was made using the cloud catalogue of \citet{Roman-Duval+09}, and an association between a CHIMPS clump and a GRS cloud was made if the CHIMPS clump centroid falls within half of the FWHM-extent of the GRS cloud about the cloud's centre. In cases where a CHIMPS clump matches with multiple GRS clouds, the association with the smallest velocity difference was preferred. 
    \item \addtocounter{enumi}{2} The final step was to make associations between the remaining CHIMPS sources with undetermined distances using a final volumetric search. An ellipsoidal volume of $0 \fdg 3 \times 0 \fdg 3 \times 10\,$\kms\ was searched around the $\ell, b, v$ centroid of each remaining CHIMPS clump in order to identify and make an association with another CHIMPS clump centroid with an existing distance assignment. The choice of the search volume follows the tolerance determined to be appropriate for friends-of-friends grouping by \citet{Wienen+15}, and corresponds to the median angular size and maximum linewidth of molecular clouds found by \citet{Roman-Duval+09}. Where an association within the search volume could be made, the same solution to the kinematic distance ambiguity was adopted for the previously unassigned clump, and the kinematic distance corresponding to the newly assigned clump's coordinates was chosen. In cases where multiple CHIMPS clumps with distance assignments are located within a particular search volume, the closest match -- in terms of the length of normalised connecting vector -- was preferred. Sources that have a poor reliability flag (with a value of 1: see Sect.~\ref{sec:sourceextraction}) were disqualified from providing a secondary distance match in this step.
\end{enumerate}

\begin{table}
  \centering
  \caption{A summary of the number of kinematic distance solutions identified by each of the methods outlined in Sect.~\ref{sec:distances}.}
  \label{tab:distancesummary}
  \resizebox{\hsize}{!}{%
  \begin{tabular}{rrl}
    \hline \hline
    \noalign{\smallskip}
    Assignment & No. distances & Reference\\
    method 	   & assigned	   & catalogue\\
    \hline
    \noalign{\smallskip}
    i) 		& 52 	& --	\\
    ii) 	& 306 	& --	\\
    iii) 	& 585 	& \citet{Urquhart+18}	\\
    iv) 	& 7 	& \citet{Wienen+15}	\\
    v) 		& 14 	& \citet{Lumsden+13} \\
    vi) 	& 438 	& \citet{Ellsworth-Bowers+15a} \\
    vii) 	& 88 	& \citet{Eden+12}, \\
    & & \citet{Eden+13} \\
    viii) 	& 401 	& \citet{Rathborne+09}, \\
    & & \citet{Roman-Duval+09}\\
    ix) 	& 493 	& \citet{Roman-Duval+09} \\
    x) 		& 2426 	& This work	\\
    \noalign{\smallskip}
    \hline
    \noalign{\smallskip}
    Total & 4810 & \\
    Unassigned & 189 & \\
    \noalign{\smallskip}
    \hline
    \end{tabular}}
\end{table}

\begin{figure}
  \centering
  \resizebox{\hsize}{!}{\includegraphics{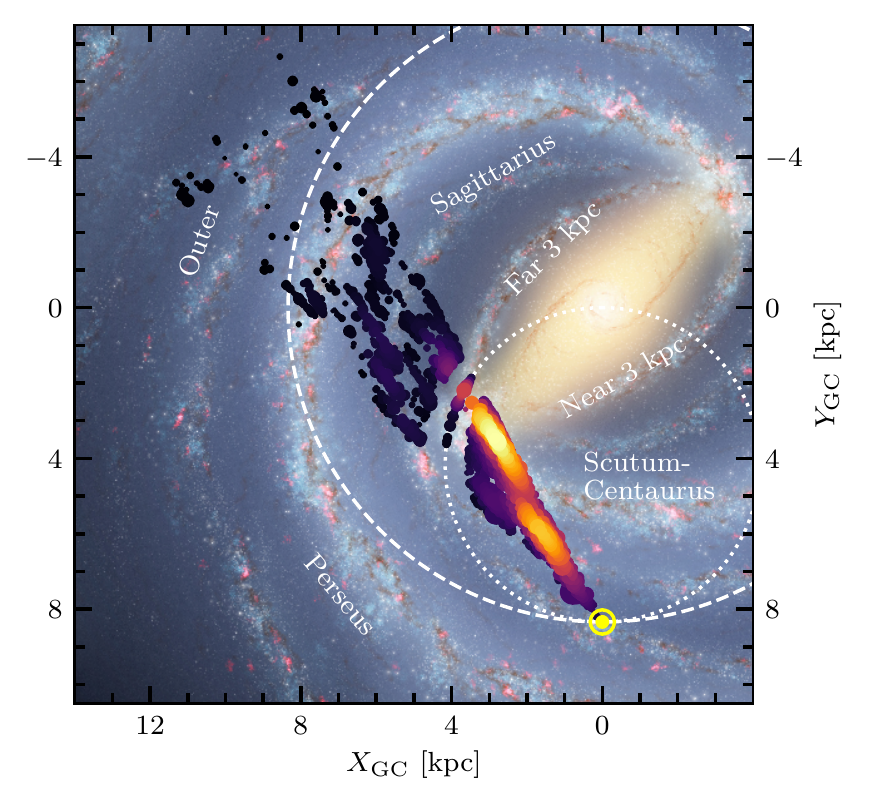}}
  \vspace{-3mm}
  \caption{A top-down view of the distribution of the 4810 $^{13}$CO (3--2) sources with kinematic heliocentric distances derived from the \citet{Reid+14} rotation curve. The relative brightness of the sources are indicated by the marker sizes, which have been normalised by the square-root of the integrated intensity after multiplying by the square of the distance. The relative source density is indicated according to the colour scale on the marked points, with black showing low density, and yellow showing high density. The background image was created by Robert Hurt of the Spitzer Science Center in consultation with Robert Benjamin at the University of Wisconsin-Whitewater and is described in \citet{Churchwell+09}. The location of the Sun has been marked, and the Solar circle and locus of the tangent points have been marked as the white dashed, and dotted lines, respectively.}
  \label{fig:distances}
  \end{figure}

\begin{SCfigure*}
  \centering
  \includegraphics[width=0.7\textwidth]{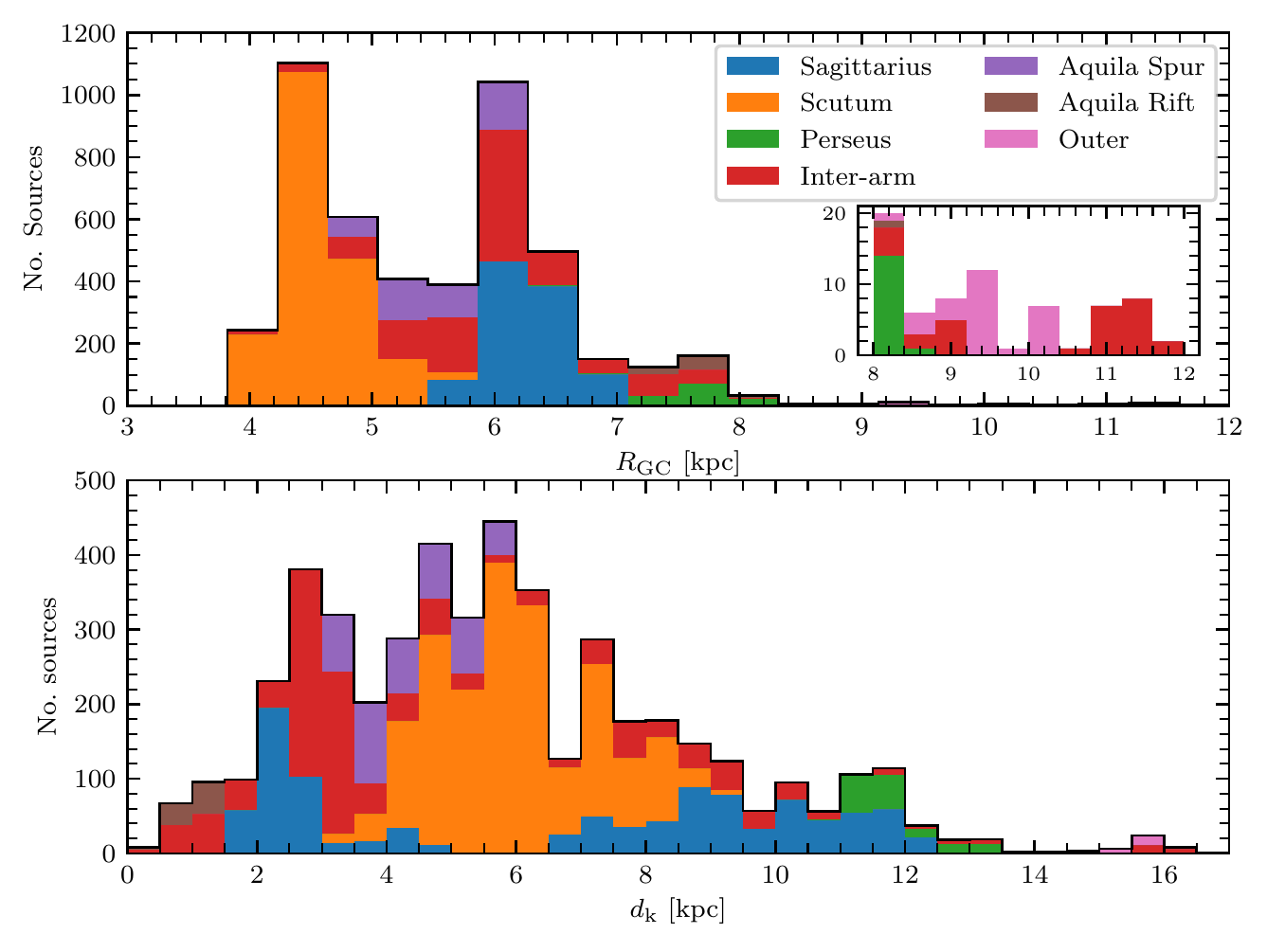}
  \vspace{-3mm}
  \caption{Distributions of the Galactocentric (top panel) and heliocentric (bottom panel) distances to the CHIMPS $^{13}$CO (3--2) sources, with spiral-arm assignments made by the Bayesian Distance Estimator of \citet{Reid+16}, which uses the models of \citet{Reid+14}. For clarity, the top panel contains an inlay showing a close-up of the distribution in the Outer Galaxy. The bin sizes used for the Galactocentric and heliocentric distributions were 400\,pc and 500\,pc, respectively.}
  \label{fig:distance_distribution}
\end{SCfigure*}

The number of distance assignments made at each of the stages outlined above is summarised in Table~\ref{tab:distancesummary}. In total, kinematic distances to 96\% of the total number of extracted sources were assigned using these methods. In Fig.~\ref{fig:distances}, the positions of the extracted $^{13}$CO (3--2) CHIMPS sources are overlaid on a top-down schematic of the Milky Way. The distributions of both Galactocentric and heliocentric distances to the sources are shown in Fig.~\ref{fig:distance_distribution}, in which we also break down the samples according to the spiral arms associated with the \citet{Reid+14} kinematic distances. The \citet{Reid+16} kinematic distance calculator assigns spiral arm associations based on the $\ell,b,v$ coordinates of a source, and its intersection with spiral arm loci of that model. The sample consists of 3120 sources lying within the four main arms, the Scutum-Centaurus, Sagittarius-Carina, Perseus, and Outer arms (while the last is also thought to be an extension of the Norma arm), while 75 and 452 sources lie within the smaller Aquila Rift and Aquila Spur features respectively. In total, 1659 of the sources with distance determinations lie outside of the four major spiral arms, and we consider these sources to reside within the inter-arm regions (which includes the Aquila Rift and Aquila Spur), and the assignments of the remaining 31 were considered to be uncertain.

We do not detect any sources located within 4\,kpc of the Galactic centre since we do not approach sufficiently central longitudes. Two large peaks in the Galactocentric distribution can be seen at $\sim 4.5$\,kpc and $\sim 6.5$\,kpc, corresponding to the Scutum and Sagittarius spiral arms, with a smaller peak at $\sim 7.5$\,kpc that is associated with the Perseus arm. In the heliocentric distribution, the peaks at $\sim 3$ and 5\, kpc correspond to the near Sagittarius and Scutum arms. Part of the Scutum arm falls at the tangential distance, and sources in this region are artificially bunched together at a distance of $\sim 7$\,kpc. This can be seen as a dearth of sources at 6.5\,kpc, which is also clearly visible in Fig.~\ref{fig:distances}, though the gap on the far side of the tangent point is not easily discernible in the lower panel of Fig.~\ref{fig:distance_distribution}, as it falls at a similar location to the start of the far Sagittarius arm, the far side of the Scutum arm, and also the location of potentially inter-arm material which creates the broad peak in the distribution at 8--10\,kpc. The peak at 12\,kpc is similarly composed of a mixture of the far Sagittarius arm and the Perseus arm, and the Outer arm is easily visible as the small peak at $\sim 16$\,kpc.

Determining distances to the large numbers of sources now being produced by Galactic plane surveys remains a significant challenge. While the `gold standard' of parallax measurements is advancing, the Bar and Spiral Structure Legacy (BeSSeL) Survey\footnote{\url{http://bessel.vlbi-astrometry.org}} and the Japanese VLBI Exploration of Radio Astrometry (VERA)\footnote{\url{http://veraserver.mtk.nao.ac.jp}} are currently limited to the relatively sparse sampling of high-mass star-forming regions containing masers (methanol and water masers are typically used), and while \textit{Gaia} will revolutionise this field for relatively unobscured, nearby or massive star-forming regions, many of the intermediate and low-mass or very young star-formation sites in the Galactic plane will remain unprobed. While we do make associations between sources in order to distinguish between the kinematic distance solutions for individual sources, we do not assign common distances to these groups as has been done in, for example, the friends-of-friends analysis used for ATLASGAL clumps \citep[e.g.,][]{Wienen+15,Urquhart+18}. The effect of this can be seen in various places of Fig.~\ref{fig:distances}, in which groups of objects that are probably located at the same distance are smeared out along the line of sight as any peculiar motions along the line of sight are interpreted as real differences in distance. 

The uncertainties adopted for these kinematic distances were those determined by the probability related to the kinematic distance estimates alone in the \citet{Reid+16} Bayesian distance formalism. The \citet{Reid+16} method assumes that line-of-sight velocities have a random dispersion of 5\,\kms, as a consequence of virial motions induced by a typical mass of 10$^6$\,M$_\sun$ within $\sim$100\,pc. However, it is important to note that the distance uncertainties quoted here do not account for the uncertainty between the near- and far-kinematic distance solutions (i.e. incorrect assignments).

The ordered approach of Methods i)--x) listed in this section naturally assumes a hierarchy of reliability, with methods deemed more reliable used first in order that samples using the least reliable methods can be excluded from future studies without the need to recalculate any dependent distances. For instance, any erroneous assignment of a kinematic distance solution in Steps i)--ix) could form the seed for many further incorrect distance assignments in Step x). The order was chosen to reflect both the types of sources that have been considered, and generally placing a greater emphasis on sources that have been analysed by eye rather than automated methods. To check for self-consistency, we repeated the volumetric search Stages iii)--vii) after combining all of the individually used catalogues into a single catalogue, to repeat the search in a more bias-free manner. After propagating this through the remaining Stages viii)--x), there is a 90\% agreement of kinematic distance solution assignments, showing that the exact order of the volumetric searches does not offer a dominant source of bias.

\section{Clump properties} \label{sec:properties}

The sources that have been extracted from the CHIMPS data broadly cover the range of parameter space in mass, size and density that are usually ascribed as `clumps' in the literature \citep[e.g.][]{Bergin+Tafalla07}, and so we adopt this term hereafter for the sake of convenience. The {\sc FellWalker} masks (described in Sect.~\ref{sec:sourceextraction}) were used to extract the optical depths, excitation temperatures, and $^{13}$CO column densities from the cubes generated as in Sect.~\ref{sec:LTE}, while the source sizes were determined from the native 15.2-arcsec-resolution $^{13}$CO (3--2) data.

The shapes of molecular clouds are complex, and the {\sc FellWalker} source extraction reports the intensity-weighted rms deviation of voxels from the centroid in the orthogonal $\ell$, $b$ and $v$ axes \citep[see][for more details]{Berry15}, as opposed to any elliptical fitting. For sources with purely Gaussian profiles, these rms sizes would return the standard deviation of the profile in the corresponding axis, which may be converted into FWHM by multiplying by a factor of $\sqrt{8 \ln 2}$. The reported sizes are a convolution of the underlying source size with the 15.2 arcsec telescope beam, and so the reported sizes in the $\ell$ and $b$ axes were deconvolved to remove these effects. Although the intensity-weighted rms sizes are not strictly standard deviations because the sources are not all perfectly Gaussian, this size deconvolution only makes a significant change to the reported source size for objects which are only slightly larger than the beam size, and such objects generally are compact and Gaussian-like. The catalogued peak intensity values are also modified by these reported smoothing effects, and are rescaled as stated in \citet{Berry15}.

In this section, we define our `full sample' as the 3553 sources which have the highest reliability flag, and which have determined masses (i.e. both a distance and column density determination). We note that there are a total of eight sources that have good reliability flags and defined distances, but no column density determination due to their position lying outside of the COHRS survey latitude range. It is important to consider the effect of distance biases upon the derived clump properties and so in this section we also define a `distance-limited sample' of clumps, against which we will compare any relationships between physical quantities of the full sample. This sample consists of the 671 clumps with distances in the range $8 \leq d_\mathrm{k} \leq 12$\,kpc, with a good reliability flag and with determined column densities. The distance range was chosen so that the spatial resolution element between the nearest and most distant sources differs by no more than 50\%, while covering a significant fraction of the full sample.

In this section, we adopt an uncertainty on the one-dimensional linewidth of:

\begin{equation}
    \Delta\,\sigma_v = \left( \frac{13.3}{\mathrm{S/N} + 5.5}\right) \mathrm{km\,s}^{-1},
\end{equation}

\noindent where S/N is the peak signal-to-noise ratio. This uncertainty was derived empirically through injecting synthesised Gaussian sources into a sample of the data, and comparing the recovered linewidths with the input. The recovered linewidth tends to the input linewidth as the S/N increases, but can be undefined when below $\sim50\%$ of the channel width.

\begin{figure*}[ht!]
	\centering
	\includegraphics{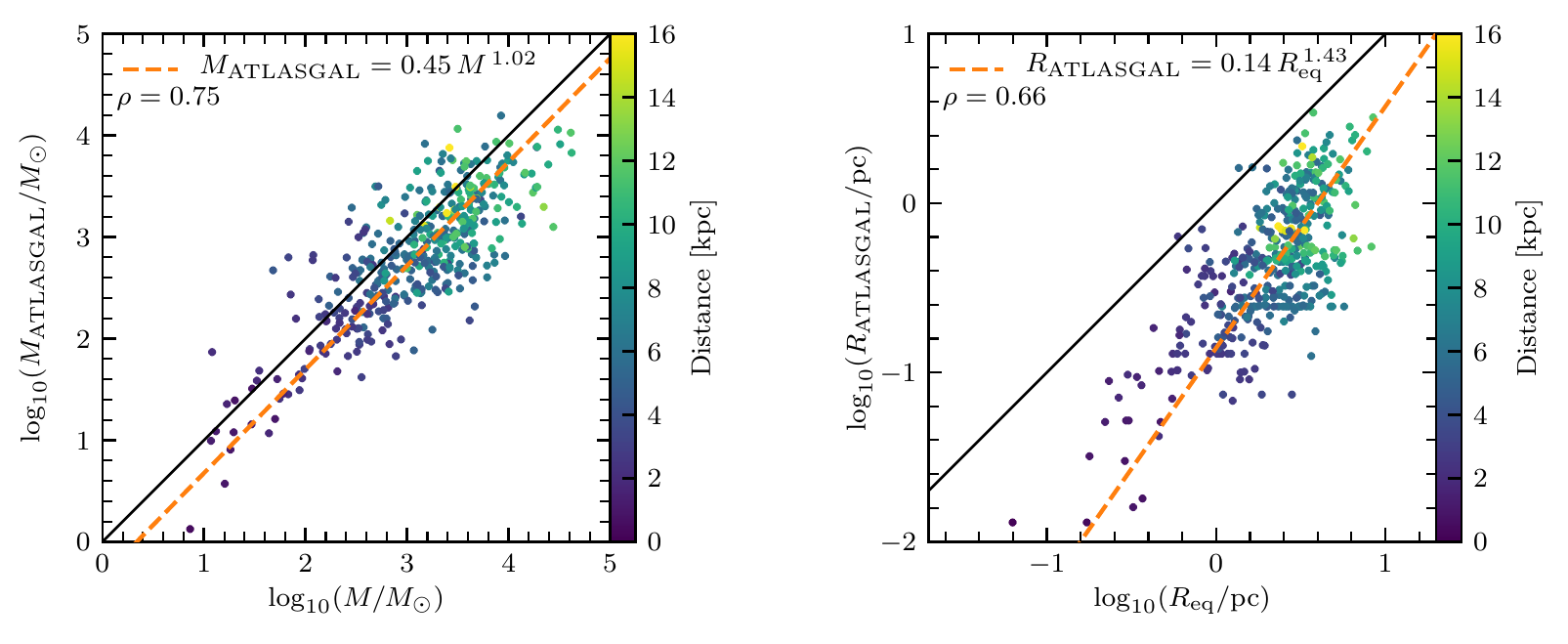}
	\vspace{-4mm}
	\caption{A comparison of the masses (left panel) and radii (right panel) of a sample of 412 CHIMPS clumps that match uniquely to ATLASGAL clumps. The ATLASGAL masses and radii have been rescaled to match the distance assigned to their corresponding CHIMPS association, and the colour of each point shows the source distance. The solid black line shows the 1:1 relationship, and the dashed orange line is the fitted relationship, and the Spearman's rank correlation coefficient is shown.}
	\label{fig:CHIMPSvsATLASGAL}
\end{figure*}

\subsection{Basic physical properties} \label{sec:masses}

For each clump, the total mass is determined from the column density integrated over all its constituent $\ell$, $b$, $v$ voxels using the following formula:

\begin{equation} \label{eq:mass} 
	M = \mu \, m_\mathrm{p} \, R_{13}^{-1} \, \,  d_\mathrm{k}^2 \sum\limits_{\ell bv} N_{13}(\mathrm{total})_{\ell bv},
\end{equation}

\noindent where $\mu$ is the mean mass per H$_2$ molecule, taken to be 2.72, accounting for a helium fraction of 0.25 \citep{Allen73}, $m_\mathrm{p}$ is the mass of a proton, $R_{13}^{-1}$ is the abundance ratio of H$_2$ compared with $^{13}$CO, and $d_\mathrm{k}$ is the clump's kinematic distance. The conversion  $R_{13}^{-1}$ is calculated in two steps, and we adopt a ratio of $R_{12}/R_{13}$ that varies as a function of Galactocentric distance as prescribed by \citet{Milam+05},  and a value of $R_{12} = 8.5 \times 10^{-5}$ \citet{Frerking+82} is adopted for all sources, with an uncertainty taken to be 30\%. $R_{12}/R_{13}$ has a value of approximately 50 for the median Galactocentric distance of 5.5\,kpc within the sample.

To compare the masses derived for the CHIMPS clumps with an independent measure, their ATLASGAL dust continuum counterparts were identified. Unique associations between the most robust CHIMPS clumps (with a reliability flag value of 3) and ATLASGAL clumps \citep[from the catalogue of][]{Urquhart+18} were made by searching in a volume with a radius of three CHIMPS resolution elements (i.e. 45 arcsec in $\ell$ and $b$, and 1.5\,\kms\ in velocity) around the position of the peak $^{13}$CO (3--2) intensity. 426 unique CHIMPS-ATLASGAL associations were made in this way, of which 412 have distance determinations in both catalogues, and their properties are compared in the left panel of Fig.~\ref{fig:CHIMPSvsATLASGAL}. The ATLASGAL masses have been rescaled to adopt the distance of the corresponding CHIMPS association in order to reduce the level of scatter resulting from only differences in the distance determination. Using the SciPy implementation of orthogonal distance regression (ODR; \citealt{Boggs+Rogers90}), in order to take uncertainties on both axes into account, we find that the masses between the two tracers are strongly and approximately linearly correlated (with a Spearman correlation coefficient of $\rho = 0.75$), when adjusted for distance, and the ATLASGAL clumps are found to have a mass of $\sim 45 \pm 13\%$ of the corresponding CHIMPS clump. There is no significant systematic variation in the power-law index when considering distance-limited subsets.

We define the radius of each CHIMPS clump in two ways. The first method defines the radius associated with the geometric mean of the sizes in the $l$ and $b$ axes as reported by {\sc FellWalker}:
\begin{equation} \label{eq:rsig} 
	R_\sigma = d_\mathrm{k}\sqrt{\sigma_\ell \, \sigma_b} ,
\end{equation}

\noindent where $\sigma_\ell$ and $\sigma_b$ are the intensity-weighted rms deviations in the $l$ and $b$ axes, deconvolved to account for the beam, and $d_\mathrm{k}$ is the assigned kinematic distance. Secondly, we define the equivalent radius of a circle with the same projected area, $A$, as the source:
\begin{equation} \label{eq:req} 
	R_\mathrm{eq} = d_\mathrm{k} \sqrt{A / \pi} .
\end{equation}

\noindent The radius given by $R_\sigma$ takes the source $^{13}$CO (3--2) emission profile into account while $R_\mathrm{eq}$ has no dependence on the emission profile. We adopt a version of $R_\sigma$ scaled by a factor $\eta$ that accounts for an average emission profile for the determinations of most of the radius-dependent parameters in this section. Since we are primarily interested in the densest regions of the clumps, where star formation is more likely to be located (assuming most clumps are centrally concentrated), and given the variations in the noise level across the survey, $R_\sigma$ will provide a more consistent measurement than $R_\mathrm{eq}$. The commonly-used conversion between the two radii of $R_\mathrm{eq} = \eta R_\sigma$ with $\eta = 1.9$ \citep[e.g.][]{Solomon+87,Rosolowsky+Leroy06,Colombo+19} agrees well with the distributions in these data, for which we find a median value of $\eta = 2.0$. We use this value to modify $R_\sigma$ where necessary to best compare with the
 literature.

In Fig.~\ref{fig:CHIMPSvsATLASGAL} we also compare the radii of the matched CHIMPS--ATLASGAL sample, and find a power-law relationship in which $R_\mathrm{ATLASGAL} = 0.14 \, {R_\mathrm{eq}^{1.4}}$. CHIMPS finds larger radii for almost all ATLASGAL clumps (after normalising to the same distances), with the largest difference at low masses. These low-mass clumps also tend to have the lowest peak column densities within the ATLASGAL catalogue, and this radius relationship might therefore be explained by the higher sensitivity of the CHIMPS survey, an attribute that we discuss in more detail in Sect.~\ref{sec:size+density}.

\begin{figure}
	\centering
	\resizebox{\hsize}{!}{\includegraphics{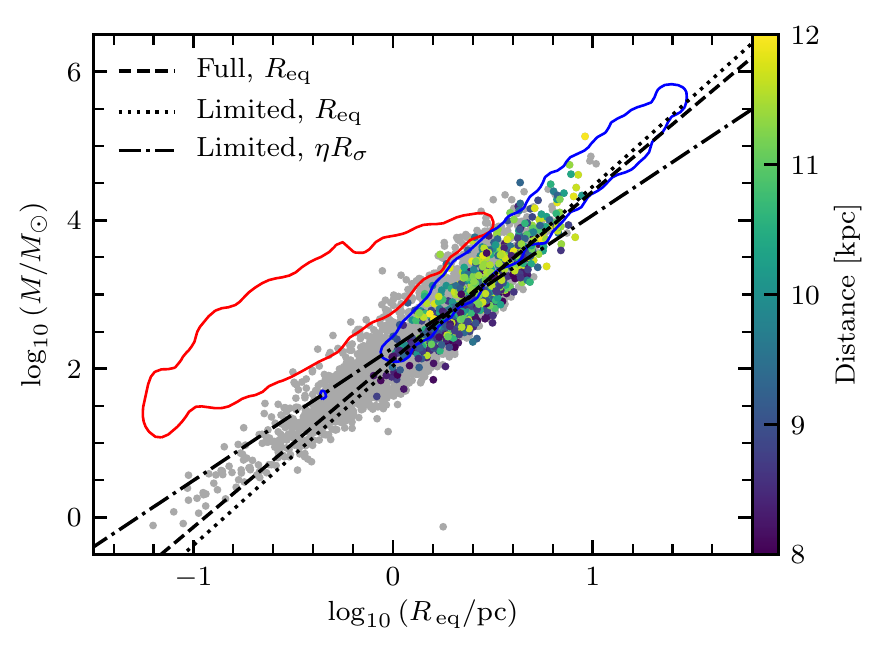}}
	\vspace{-5mm}
	\caption{The mass--radius relationship for the CHIMPS clumps (coloured points) and, for comparison, we overlay contours containing 90\% of the points in ATLASGAL \citep{Urquhart+18} in red, and GRS \citep{Roman-Duval+10} in blue. The coloured points are the distance-limited CHIMPS sample, while the grey points show the full sample. The dashed, dotted, and dot-dashed lines show the best-fit power-law to the various samples. The derived power-law fits are and Spearman's rank correlation coefficients are listed in Table~\ref{tab:M-R}.}
	\label{fig:M-R}
\end{figure}

The mass-radius (M--R) relationship is displayed in Fig.~\ref{fig:M-R} alongside the GRS molecular clouds and ATLASGAL clumps. We perform power-law fits of the M--R relationship for the full sample of clumps, and the distance-limited sample using the $R_\mathrm{eq}$ measurements, and we also plot the distance-limited sample in terms of a scaled $R_\sigma$. We plot the CHIMPS sample in terms of $R_\mathrm{eq}$ for direct comparison with the GRS sample. We note that since the radii in the ATLASGAL sample are given as $R_\mathrm{eq} = \eta R_\sigma$ with $\eta = 2.4$, we also fit a power law to the scaled $R_\sigma$ CHIMPS radii, adopting the median value of $\eta = 2.0$. The equations of the power-law fits and their correlation coefficients can be found in Table~\ref{tab:M-R}.
 
 A power-law fit using ODR finds that the relationship can be described as $M \propto R_\mathrm{eq}^{2.26 }$ for the full sample, and $M \propto R_\mathrm{eq}^{2.42}$ for the distance-limited sample. By comparison, molecular clouds in the GRS have a similar power-law exponent, for which \citet{Roman-Duval+10} found $M \propto R^{2.36}$, and dense clumps in ATLASGAL are found by \citet{Urquhart+18} to follow a shallower relationship, with $M \propto R^{1.65}$. The scatter on the CHIMPS data is much larger than that on the GRS, and probably relates to the large difference in resolution, and it is comparable to the scatter in the ATLASGAL data, which were extracted at similar resolution ($\sim 20$ arcsec). We note that, while the power-law index in $M$--$R_\mathrm{eq}$ for the full CHIMPS sample is similar to that of the GRS, the index of the power-law  in the $M$--$\eta R_\sigma$ relationship is intermediate between the ATLASGAL and GRS indices. We point out here that the choice of radius can make a significant difference in these kinds of results.

\begin{table}
    \centering
    \caption{Parameters for the ODR fits to the various mass--radius relationships shown in Fig.~\ref{fig:M-R}, including the Spearman correlation coefficients, $\rho$. Note that we adopt the median value of $\eta = 2.0$ derived for the CHIMPS sources.}
    
    \begin{tabular}{llc}
        \hline \hline
        \noalign{\smallskip}
         Sample & Relationship & $\rho$ \\
         \noalign{\smallskip}
         \hline
         \noalign{\smallskip}
         Full &  $M = (134 \pm 2) \, R_\mathrm{eq}^{(2.26 \pm 0.02)}$ & 0.92 \\
         Distance-limited &   $M = (109 \pm 6) \, R_\mathrm{eq}^{(2.42 \pm 0.05)}$ & 0.87 \\
         Distance-limited & $M = (191 \pm 13) \, \eta \, R_\sigma^{(1.79 \pm 0.06)}$ & 0.87 \\
         \noalign{\smallskip}
         \hline
    \end{tabular}
    \label{tab:M-R}
\end{table}

We also calculate the mean (volumetric) particle density measured over the full extent of the clump by:
\begin{equation}
	\bar{n}(\text{H}_2) = \frac{3}{4\pi} \frac{M}{\mu m_\mathrm{p}  R_\mathrm{eq}^{\,3}}.
\end{equation}

The distributions of clump masses, radii and average volume densities are shown in Panels a), b) and c) of Fig.~\ref{fig:propertydistributions}, and are compared with the corresponding distributions from the GRS molecular clouds \citep{Roman-Duval+10} and ATLASGAL clumps \citep{Urquhart+18}, which have been restricted to sources lying within the CHIMPS survey area. Alongside the 3553 CHIMPS clumps from the full sample, the distributions contain a total of 911 ATLASGAL clumps and 259 GRS molecular clouds. Although we refer to the sources extracted from the CHIMPS data as `clumps' in this paper due to the correspondence between their derived masses and radii and the values of objects described as `clumps' across the literature, the term `cloud fragments' might also equally be applied. We do not make any attempt to fit a power-law to the mass function of the CHIMPS clumps because there is no single completeness limit in these data (see Appendix~\ref{app:completeness}). The turnover of the distribution is often attributed to the completeness limit of the data -- that is, the mass limit below which sources are not dependably extracted (and therefore not fitted in any power-law) -- but this limit depends on the size in both spatial and spectral axes, the local noise level, and the source density profile in addition to the total mass.

\subsection{Dynamic state} \label{sec:dynamic_state}

\begin{figure*}[ht]
	\centering
	\includegraphics[width=17cm]{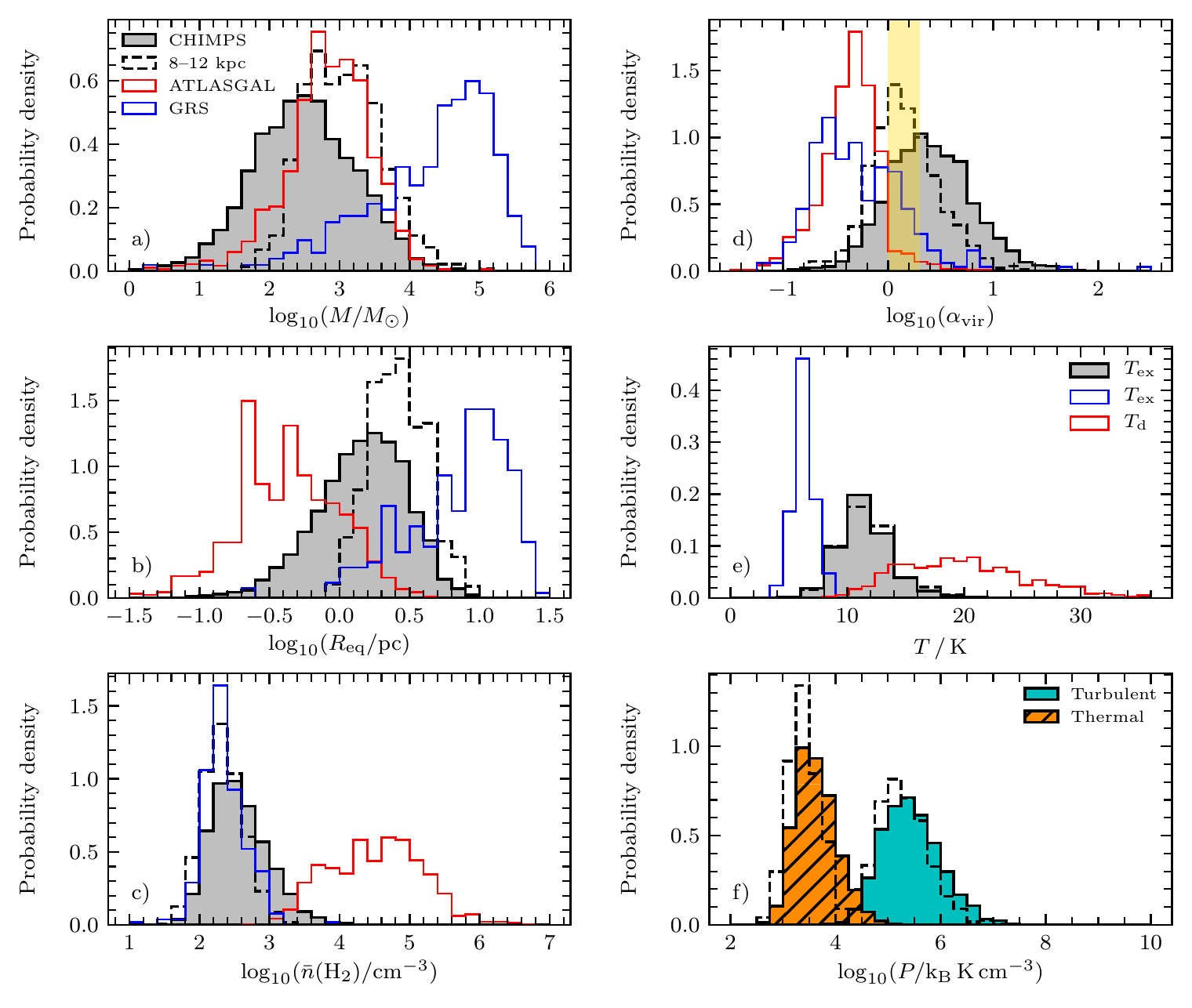}
    \caption{Panels a), b), c) and d) display the distributions of total mass, equivalent radius, average number density, and virial parameters  of the CHIMPS clumps (black histogram), GRS molecular clouds (blue) and ATLASGAL clumps (red). Panel e) shows the mean excitation temperature distributions for the CHIMPS clumps and GRS molecular clouds, alongside the dust temperature distribution of ATLASGAL clumps \citep{Urquhart+18}, adopting the same colour coding as the preceding panels. In addition, Panel f) displays the thermal and turbulent pressures for the CHIMPS clumps. The yellow shaded region in Panel d) shows the expected values for clumps in approximate virial equilibrium. In each panel, the empty dashed-outline histograms show the CHIMPS clump distributions for the corresponding property, but for the distance-limited sample with $8 \leq d_\mathrm{k} \leq 12$\,kpc.}
    \label{fig:propertydistributions}
\end{figure*}

The dynamic state of the molecular clumps -- whether they are expanding, collapsing or in some quasi-stable equilibrium -- can be assessed by using the virial theorem (when twice the kinetic energy is equal to the gravitational energy, $2K +\Omega =0$). The virial parameter, the ratio of a cloud's spherically symmetric virial mass to its total mass, is defined as:
\begin{equation}
	\avir =\frac{3 \sigma_v^2  \eta R_\sigma}{GM},
\end{equation}
\noindent where $\sigma_v$ is the one-dimensional velocity dispersion of a clump with radius $R_\sigma$ and mass $M$, and $G$ is the gravitational constant. Here we follow the \citet{MacLaren+88} prescription of the virial parameter, and assume a $\rho(r) \propto 1/r^2$ spherical radial density distribution in order to compare directly with the GRS molecular clouds \citep{Roman-Duval+10}, and modify $R_\sigma$ with the multiplicative factor $\eta = 2.0$ to account for the median emission profile.  A value of $\avir=1$ indicates virial equilibrium (in the absence of significant magnetic fields), while $\avir = 2$ describes a clump with an equipartition of gravitational and kinetic energy. When $\avir < 1$, the system is unstable to gravity and is collapsing, in the absence of other supporting pressures, whereas $\avir > 2$ suggests that it is dissipating, as its kinetic energy dominates its gravitational energy, and $\avir \sim 1-2$ describes a clump that is in approximate equilibrium. However, it has also been argued \citep[e.g.][]{Kauffmann+13} that cloud fragments undergoing collapse will tend to be found with $\avir \sim 2$, as the rapid infall of gas serves to increase velocity dispersions in response. \citet{Kauffmann+13} argue that cloud fragments with $\avir \ll 2$ are more likely to be either supported by significant magnetic fields, or may indicate ongoing high-mass star formation.

We adopt the scaled intensity-weighted radius, $\eta R_\sigma$, in this parameterisation of the virial parameter in order to best compare with both the GRS \citep{Roman-Duval+10} and ATLASGAL \citep{Urquhart+18}, which use the equivalent radius, and a scaled intensity-weighted rms radius, respectively. The use of either $R_\sigma$ or $R_\mathrm{eq}$ alone in this equation both have their advantages and disadvantages, and a weighted $R_\sigma$ represents the best compromise. The intensity-weighted radius we adopt allows for a greater weighting in gravitational energy to be assigned to the more-concentrated regions of the mass distributions whilst simultaneously providing a measurement that is less S/N dependent (which is important when considering the noise level variation across the CHIMPS survey). The scaling factor compensates for the mass located at larger radii that would otherwise be interpreted as being located within $R_\sigma$, which would result in an overestimate of the gravitational energy. The drawback of this method is that it effectively converts $R_\sigma$ to $R_\mathrm{eq}$ assuming that all clumps follow a density profile that is the average of all reliable clumps across the survey. There is an intrinsic uncertainty of a factor of two on all virial parameter measurements, due to both modelling the sources as spherically symmetric and characterised by a single radius, as well as the particular radial profile shown, and so we caution against interpreting these measurements as definitive assessments of the sources' dynamic status.

The distributions of the virial parameters are shown in Panel d) of Fig.~\ref{fig:propertydistributions}. The distribution of the virial parameters is significantly skewed, and we find a median value of $\avir = 2.4^{+2.1}_{-1.1}$, where the quoted uncertainties show the range of the first and third quartiles, and the distribution features an extended tail with $\avir > 10$. This would seem to suggest that the clumps traced by $^{13}$CO (3--2) do not represent a single phase, but cover fragments of molecular clouds which are both dispersing and collapsing, while the majority appear to be approximately in virial equilibrium. The clump masses measured by CHIMPS do not describe the full picture since the (3--2) transition, with a critical density of $\sim 10^4\,$\pcmmm, is mostly sensitive to the relatively dense gas.

These clumps sit in a wider gravitational potential caused by all of the lower-density molecular gas that CHIMPS does not trace, in addition to the emission that is missed due to finite sensitivity. While we know that both the detection threshold used in the source extraction method and the finite sensitivity of the data can result in CHIMPS clump masses being reported as systematically light, the comparison of the CHIMPS and ATLASGAL clumps in Fig.~\ref{fig:CHIMPSvsATLASGAL} shows that the $^{13}$CO masses are generally larger than their dust-clump counterparts, indicating that mass underestimates are unlikely to be a significant problem. Further, the `missing mass' in low-density gas is also likely to be located at larger radii, so the effect that this missed material in our mass estimate would have on virial parameters is probably negated by the smaller radii reported for the CHIMPS emission compared with what might be seen in (1--0) with similar sensitivity and resolution. The extended tail of the distribution that reaches high values of $\avir$ ($\lesssim 100$), suggests that there are some clumps that are either pressure-confined or are, as seems more likely, transient structures with respect to their dynamical crossing timescales $2 R_\mathrm{eq}/\sigma_v$.

The mean excitation-temperature distribution (shown in Panel e) of Fig.~\ref{fig:propertydistributions}) has a median value of $T_\mathrm{ex} = 11.3^{+1.4}_{-1.2}$\,K, which is slightly larger than the mean excitation-temperature distribution reported for molecular clouds in the GRS \citep{Rathborne+09, Roman-Duval+10} to which they are most comparable. All the distributions exhibit a very sharp lower limit, with almost no clumps exhibiting excitation temperatures lower than approximately 6.0\,K, which is most likely a selection effect caused by the limited sensitivity of $^{13}$CO (3--2) to colder gas. The mean temperatures of sources of both of these molecular tracers are considerably cooler than the dust temperatures found in the dense clumps traced by ATLASGAL.

We also calculate the turbulent pressure, which can be determined according to:
\begin{equation} \label{eq:Pturb} 
	P_\mathrm{turb}/\,k_\mathrm{B} = \mu m_\mathrm{p} \bar{n}(\text{H}_2) \, \sigma_\mathrm{NT}^{\,2}/ \, k_\mathrm{B},
\end{equation}
\noindent which has units of K\,\pcmmm, where $k_B$ is Boltzmann's constant and $\sigma_\mathrm{NT}$ is the non-thermal component of the three-dimensional velocity dispersion, where $\sigma_\mathrm{NT}^{2} = 3\sigma_v^2 - k_\mathrm{B}T_\mathrm{ex}/m_{^{13}\mathrm{CO}}$ and $m_{^{13}\mathrm{CO}}$ is the mass of a $^{13}$CO molecule. The turbulent pressure distribution (shown in Panel f) of Fig.~\ref{fig:propertydistributions}) has a mean value of $P_\mathrm{turb} / k_\mathrm{B} = 2.5 \times 10^5\,$K\,\pcmmm\ and a standard deviation of 0.4 dex. For reference, the total mid-plane pressure in the solar neighbourhood has a value of  $P_\mathrm{turb}/k_\mathrm{B} \sim 10^5\,$K\,\pcmmm, while $P_\mathrm{turb}/k_\mathrm{B} \sim 10^9\,$K\,\pcmmm\ may be found in the Galactic centre \citep[e.g.][]{Rathborne+14}. The range of turbulent pressures spanned is consistent with these numbers, since the majority of the clumps lie inside the solar circle, and we do not probe within 4\,kpc of the Galactic centre. We also find that the turbulent pressures are $\sim 60$ times greater than the corresponding thermal pressures.

\subsection{Clump physical-properties catalogue}

\begin{table*}
	\centering
	\caption{Physical properties derived for the ten CHIMPS $^{13}$CO (3--2) clumps with the greatest integrated emission.}
	\label{tab:properties}
    \resizebox{\hsize}{!}{
    \begingroup
    \renewcommand{\arraystretch}{1.2} 
    \begin{tabular}{lcccccccccccc}
    \hline \hline
    \noalign{\smallskip}
Designation & KDA & Method & $d_\mathrm{k}$ & $R_\mathrm{GC}$ & $M$ & $R_\sigma$ & $R_\mathrm{eq}$ & $\bar{n}(\mathrm{H}_2)$ & $\alpha_\mathrm{vir}$ & $\bar{T}_\mathrm{ex}$ & $P_\mathrm{therm} / k_\mathrm{B}$ & $P_\mathrm{turb} / k_\mathrm{B}$ \\
 &  &  & [kpc] & [kpc] & [$10^3 M_\sun$] & [pc] & [pc] & [$10^3$\,cm$^{-3}$] &  & [K] & [$10^4$\,K\,cm$^{-3}$] & [$10^6$\,K\,cm$^{-3}$] \\
  (1) & (2) & (3) & (4) & (5) & (6) & (7) & (8) & (9) & (10) & (11) & (12) & (13)  \\ 
 \noalign{\smallskip}
 \hline
 \noalign{\smallskip}
G043.167+00.017 & F & iii & 11.96 & 8.19 & 135 & 3.55 & 9.16 & 0.62 & 0.86 & 22.1 & 1.4 & 14.3 \\
G034.238+00.115 & N & iii & 3.43 & 5.83 & 13 & 1.19 & 2.76 & 2.22 & 0.58 & 14.7 & 3.3 & 10.2 \\
G034.248+00.166 & N & iii & 3.38 & 5.86 & 19 & 1.42 & 3.18 & 2.10 & 0.51 & 12.7 & 2.7 & 10.1 \\
G043.163-00.031 & F & iii & 11.37 & 7.78 & 56 & 3.15 & 7.66 & 0.44 & 1.18 & 18.4 & 0.8 & 6.5 \\
G029.910-00.059 & N & iii & 6.25 & 4.27 & 10 & 1.78 & 4.12 & 0.49 & 2.09 & 20.5 & 1.0 & 4.0 \\
G030.838-00.059 & N & iii & 5.91 & 4.45 & 8 & 1.77 & 4.21 & 0.40 & 3.35 & 17.3 & 0.7 & 4.5 \\
G030.722-00.098 & N & viii & 5.21 & 4.69 & 19 & 1.64 & 3.93 & 1.10 & 1.56 & 14.0 & 1.5 & 14.0 \\
G029.961-00.015 & N & iii & 5.83 & 4.39 & 8 & 1.53 & 3.49 & 0.69 & 1.64 & 20.5 & 1.4 & 4.4 \\
G030.437-00.235 & N & iii & 6.45 & 4.29 & 9 & 2.30 & 5.44 & 0.19 & 1.41 & 17.1 & 0.3 & 0.7 \\
G029.860-00.050 & N & iii & 6.05 & 4.32 & 7 & 1.59 & 3.75 & 0.48 & 1.78 & 21.2 & 1.0 & 2.8 \\
    \noalign{\smallskip}
    \hline
    \noalign{\medskip}
    \multicolumn{13}{p{1.1\textwidth}}{\textbf{Notes.} The columns detail the following: (1) source designation; (2) kinematic-distance-ambiguity (KDA) solutions, where near, tangential and far kinematic distances are designated by `N', `T' and `F', respectively; (3) method used to resolve the KDA; (4) heliocentric distance; (5) Galactocentric distance; (6) mass of H$_2$; (7) intensity-weighted rms radius; (8) equivalent radius; (9) volume-averaged particle density; (10) virial parameter; (11) mean voxel excitation temperature; (12) thermal pressure; and (13) isotropic turbulent pressure. Only a portion of the full table is shown here to illustrate its form and content. The full table can be downloaded in a machine-readable format from the CANFAR archive listed in Appendix~\ref{app:dataproducts}.}
    \end{tabular}
    \endgroup}
\end{table*}

We present a sample of the catalogue of physical properties, calculated as described in this section in Table~\ref{tab:properties}. The sources listed are the same as those in Table~\ref{tab:FWcat}, and are the ten sources with the greatest volume-integrated intensities. These sources all feature within the most-massive star-forming regions within the CHIMPS survey volume: W49, G34.3, and W43. We have made the full catalogue of clump properties available from the CANFAR archive, and it is described in Appendix~\ref{app:dataproducts}.

\subsection{Scaling relationships}

The scaling relationships between molecular-cloud properties, which are commonly known as `Larson's laws', have been the subject of a multitude of studies across the literature. The size--linewidth relationship measured by \citet{Larson81}, $\sigma_v \propto R^{0.38}$ -- spanning over a factor of 30 in size -- was originally interpreted as evidence that the interior motions of molecular clouds follow a continuum of turbulent motions inherited from the ISM at larger scales. Later studies \citep[e.g.][]{Myers+Goodman88} found a similar size--linewidth relationship, but tended to recover approximately $\sigma_v \propto R^{0.5}$. It was also found that the average particle density of molecular clouds follows $\bar{n}(\mathrm{H}_2) \propto R^{-1.10}$ which, when combined with the former relationship, implies that most molecular clouds are in approximate virial equilibrium, independent of their size. 

\begin{figure}
	\centering
	\resizebox{\hsize}{!}{
	\includegraphics{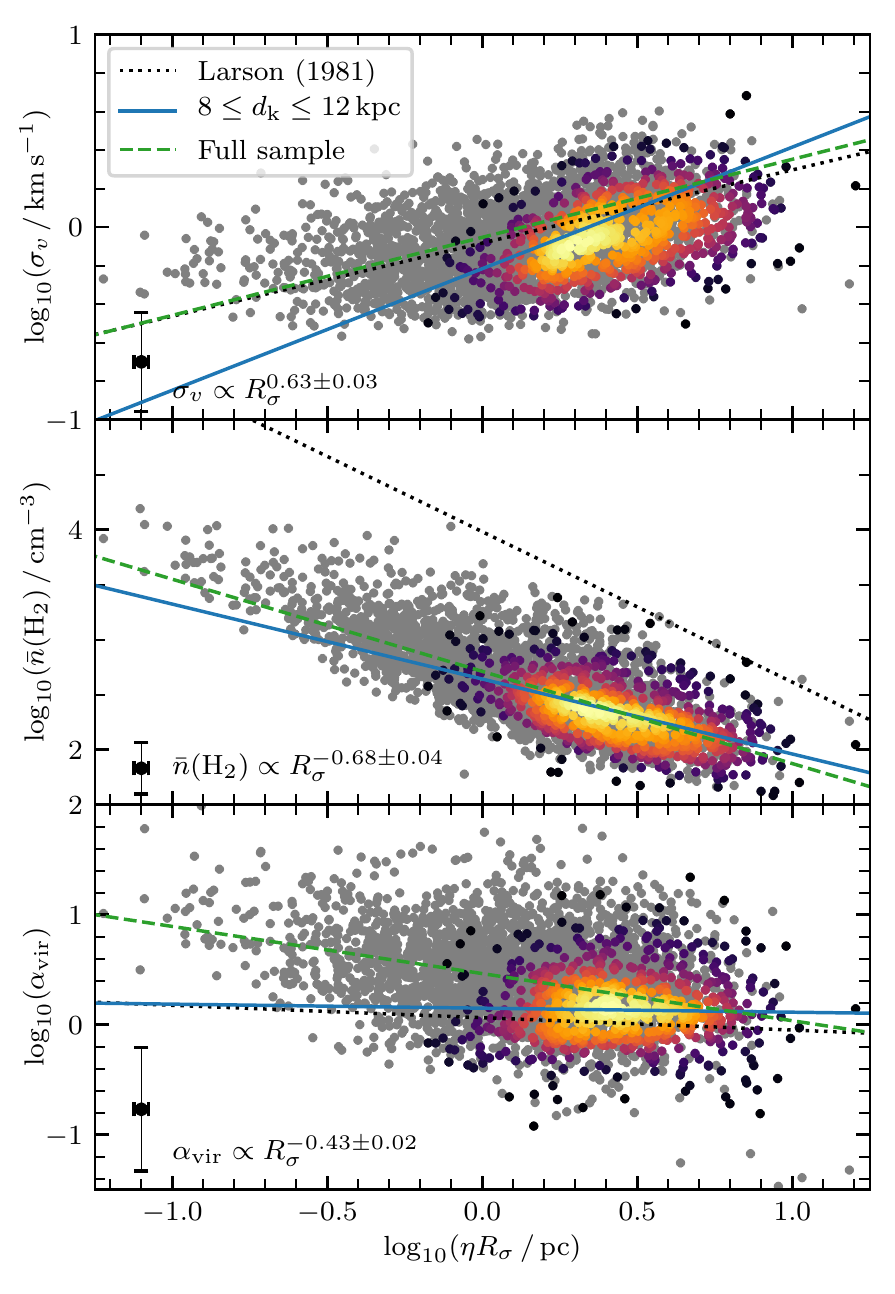}}
    \vspace{-5mm}
	\caption{The size--linewidth (top panel), size--density (middle panel), and size--virial parameter (bottom panel) relationships for the CHIMPS clumps, where the size parameter is the scaled intensity-weighted rms size (described in the text), $\eta R_\sigma$, for which $\eta = 2.0$. In each case, the grey data points show the full sample, while the distance-limited sample is shown as colour-scaled data points, for which the lightest colours indicate the highest density of points. The blue lines show the best-fitted power-law for the distance-limited sample, for which we display the median error bars on the black data point in the lower-left corner. The dashed green lines show the best fits to the full sample. The relationships derived in \citet{Larson81} are shown as black dotted lines for reference, and have been rescaled to match these quantities. Full details of the fits are given in Table~\ref{tab:scalingrelations}.}
	\label{fig:scalingrelations}
\end{figure}

The sources extracted from the CHIMPS data demonstrate a range of properties (size, mass, density) that deviate from the conventional definitions of `molecular clouds', and overlap partially with the parameter space referred to as `clumps'. It is therefore pertinent to examine whether the CHIMPS clumps agree with the molecular-cloud scaling relationships, and so we measure three of the scaling relationships for the sample of CHIMPS clumps. We primarily adopt $\eta R_\sigma$ as our size parameter, and compare the resulting best-fitting power-laws with those of \citet{Larson81} in Fig.~\ref{fig:scalingrelations}, though we do also make the comparison with $R_\mathrm{eq}$. The best-fitting relationships were determined using ODR in order to account for the uncertainties in both axes in each case, and we also compare the results for both the full sample of CHIMPS clumps (with a `good' reliability flag) and the distance-limited subset. The full list of fit parameters for each of the tested relationships can be found in Table~\ref{tab:scalingrelations}.

We find that the size--linewidth relationship for the full sample -- $\sigma_v \propto R_\sigma^{0.41}$ -- is similar to that of \citet{Larson81}, but we do see larger deviations within the distance-limited subset, for which $\sigma_v \propto R_\sigma^{0.63}$. In principle, the distance-limited subset should be impacted less by angular-resolution differences, and this distribution produces a significantly larger exponent. The discrepancy between the exponent of this size--linewidth relationship and the canonical one is even larger in the equivalent-radii case, for which we find $\sigma_v \propto R_\mathrm{eq}^{0.80}$ for the distance-limited sample. In both cases, the correlation is stronger, as measured by the Spearman's rank correlation coefficient $\rho$, for the distance-limited samples, and the $p$-values (not listed) strongly reject the null hypothesis that the correlations could have arisen by chance. The relatively high-values of the size--linewidth indices in the distance-limited samples, which ought to be more robust than those of the full samples, are closer to the values determined for high-density tracers of molecular gas found by \citet{Shetty+12} in the CMZ.

\begin{table*}
   \centering
   \caption{Parameters of the fitted power-laws for the various scaling relationships following the form $y = A (x/\mathrm{pc})^B$.}
   \begin{tabular}{@{\extracolsep{6pt}}cclrrlrr@{}}
   \hline \hline
   \noalign{\smallskip}
   Relation & Size variable & \multicolumn{3}{c}{Full sample} & \multicolumn{3}{c}{Distance-limited sample} \\ \cline{3-5} \cline{6-8}
   ($y$) & ($x$) & \multicolumn{1}{c}{A} & \multicolumn{1}{c}{B}  & \multicolumn{1}{c}{$\rho$} & \multicolumn{1}{c}{A} & \multicolumn{1}{c}{B} & \multicolumn{1}{r}{$\rho$} \\
   \noalign{\smallskip}
   \hline
   \noalign{\smallskip}
   $\sigma_v$ / \kms & $\eta R_\sigma$ & $0.89 \pm 0.07$ & $0.41 \pm 0.01$ & $0.42$ & $0.61 \pm 0.05$ & $0.63 \pm 0.03$ & $0.45$ \\
   $\bar{n}(\mathrm{H}_2)$ / \pcmmm & $\eta R_\sigma$ & $514 \pm 1$ & $-0.84 \pm 0.02$ & $-0.72$ & $438 \pm 3$ & $-0.68 \pm 0.04$ & $-0.55$ \\
   $\alpha_\mathrm{vir}$ &  $\eta R_\sigma$ & $2.91 \pm 0.05$ & $-0.43 \pm 0.02$ & $-0.28$ & $1.41 \pm 0.24$ & $-0.04 \pm 0.05$ & 	$0.04$ \\
   \noalign{\smallskip}
   \hline
   \noalign{\smallskip}

   $\sigma_v$ / \kms & $R_\mathrm{eq}$ & $0.84	\pm 0.05$ & $0.46 \pm 0.01$	& $0.47$ & $0.48 \pm 0.02$ & $0.80 \pm 0.03$ & $0.59$ \\
   $\bar{n}(\mathrm{H}_2)$ / \pcmmm & $R_\mathrm{eq}$ & $497 \pm 1$	& $-0.84 \pm 0.02$ & $-0.68$ & $415 \pm 4$ & $-0.65 \pm 0.05$ & $-0.45$\\
   $\alpha_\mathrm{vir}$ & $R_\mathrm{eq}$ & $3.10 \pm 0.05$ & $-0.55 \pm 0.02$ & $-0.37$ & $1.72 \pm 0.21$ & $-0.21 \pm 0.06$ & $-0.07$\\
    \noalign{\smallskip}
    \hline
    \noalign{\medskip}
    \multicolumn{8}{p{0.9\textwidth}}{\textbf{Notes.} The fit parameters for the power-law relationships are listed alongside the corresponding Spearman correlation coefficients. We do not explicitly list the Spearman $p$-values, which are all $< 10^{-5}$. We present the fits for both the full and distance-limited samples of CHIMPS clumps, in addition to showing the relationship as measured against both the scaled intensity-weighted rms radius, $\eta R_\sigma$, and the equivalent radius $R_\mathrm{eq}$.}
    \end{tabular}
    \label{tab:scalingrelations}
\end{table*}

The size--density relationships depart significantly from that of \citet{Larson81} in all cases, with indices significantly larger than $-1.0$, indicating that the smallest CHIMPS clumps are less dense than would be predicted by the Larson relationship. We also find a stronger negative correlation between the size and virial parameter for the full sample, with a power-law index of $-0.43$, but the relationship is close to that of \citet{Larson81} for the distance-limited sample, with an index of and $-0.04$ compared with the  $-0.14$. Although the $p$-value for the size--virial parameter relationship for the full sample is statistically significant, the relationship for the distance-limited sample is not, and the former may be due to the varying mass completeness as a function of distance.

\section{Discussion} \label{sec:discussion}
\subsection{Size and density} \label{sec:size+density}

In the CHIMPS survey area \citet{Rathborne+09} identified $\sim 260$ GRS molecular clouds with radii ranging from roughly 1 to 30 pc, associated with $\sim2300$ clumps, identified at angular resolutions of 6 arcmin and 46 arcsec, respectively, whereas there are 3664 reliable CHIMPS clumps (out of a total of 4999 extracted sources), and the majority have radii between $\sim 0.05$ and 10\,pc. The difference in angular resolution (6 arcmin for GRS clouds compared with $\sim 30$ arcsec for CHIMPS clumps) is primarily responsible for the difference in number and the sizes of sources identified, though differences are also expected due to observational selection imposed by the higher critical density of the $J$=3--2 transition.

In Fig.~\ref{fig:propertydistributions}, the normalised mass distributions for the CHIMPS and ATLASGAL clumps, and the GRS molecular clouds were compared. Although the median of the CHIMPS clump mass distribution is much lower, the CHIMPS clump distribution contains many more sources. For individual matches between the CHIMPS and ATLASGAL clump population, CHIMPS clumps tend to have a factor of two more mass, though spread over a much larger area, as Fig.~\ref{fig:CHIMPSvsATLASGAL} illustrates. The average density of CHIMPS clumps is much lower than ATLASGAL ones, though greater sensitivity of the CHIMPS survey might bias these towards lower values; if the clumps tend to be centrally concentrated, and with an extended diffuse envelope, ATLASGAL could identify the densest regions of each clump where the majority of the mass resides, but it would not be expected to recover the wider distribution of the mass. In Fig.~\ref{fig:propertydistributions} Panel c), the distribution of the mean density $\bar{n}(\mathrm{H}_2)$ of the CHIMPS clumps is compared with that of the molecular clouds found in the GRS \citep{Roman-Duval+10} and the ATLASGAL clumps \citep{Urquhart+18}. The mean density distributions for the CHIMPS clumps are much closer to the GRS distribution than to ATLASGAL, with densities slightly greater than those of the molecular clouds. The typical mean density of a CHIMPS clump is considerably lower than the critical density for the $J$=3--2 transition of CO ($\sim10^{4}$\,\pcmmm\ at a temperature of $\sim10$\,K), and may imply a volume filling factor of a few per cent if the emission is primarily thermal in origin. There is some overlap in average density between the CHIMPS and ATLASGAL clumps, which suggests that a small fraction of objects that are in an early stage of gravitational collapse are visible in both tracers.

Since the structure of molecular clouds has been found to be hierarchical \citep[e.g.][]{Blitz+Stark86, Rosolowsky+08} and possibly fractal \citep{Falgarone+91,Stutzki+98,Combes00}, it is unsurprising to find that the sources extracted from the CHIMPS data are smaller and denser than the molecular clouds of the GRS. The CHIMPS gas structures appear to be tracing some intervening density regime, covering both the molecular-cloud phase, and structures that are fragmenting to the scale of molecular clumps which contain the sites of active star formation.

\subsection{The dynamic state of the clumps}

\begin{figure}
	\centering
	\resizebox{\hsize}{!}{\includegraphics{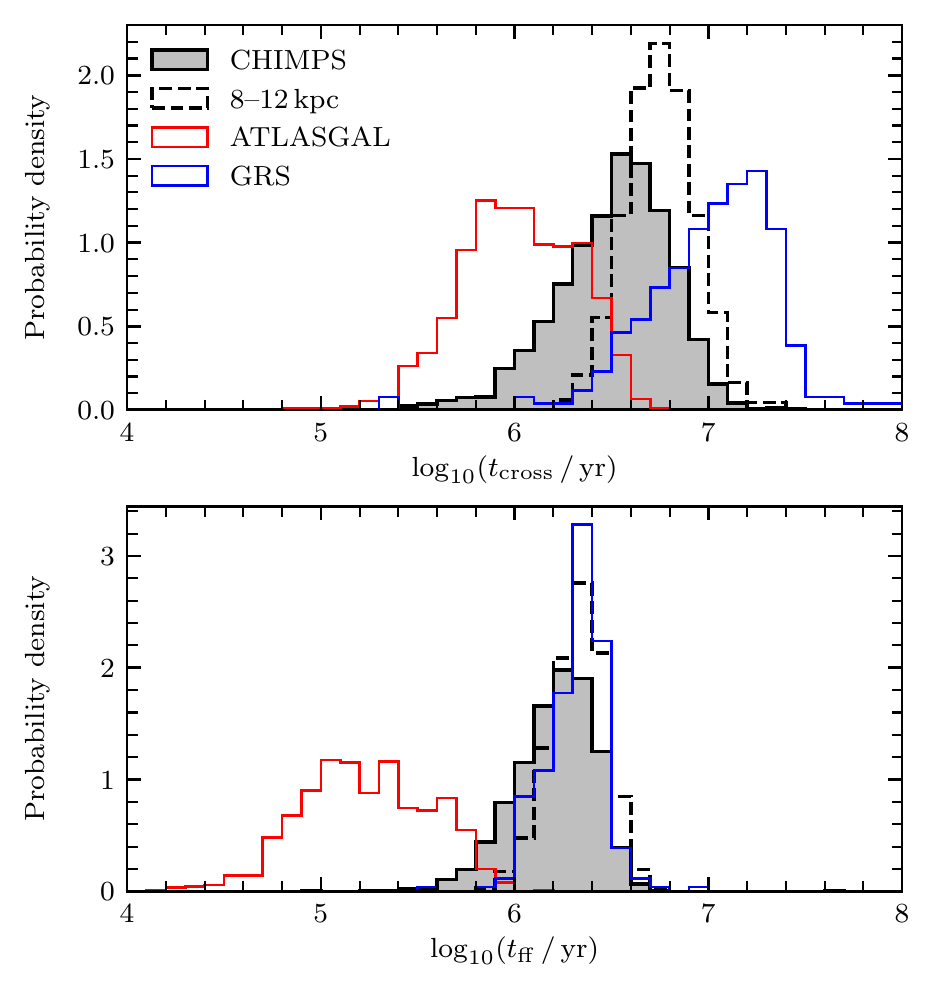}}
    \vspace{-5mm}	
	\caption{Distributions of the crossing (top panel) and free-fall (bottom panel) timescales of the CHIMPS clumps, GRS molecular clouds \citep{Roman-Duval+10}, and the ATLASGAL clumps \citep{Urquhart+18}. In both cases the bin width is 0.1 dex, and the areas have been normalised.}
	\label{fig:timescales}
\end{figure}

In Fig.~\ref{fig:propertydistributions}, the distribution of virial ratios is shown to indicate that a large fraction of the CHIMPS clumps are in approximate virial equilibrium; $\sim 80\%$ of the sample are consistent with $1 < \avir < 2$, when considering the uncertainties, which are typically on the order of a factor of 2--3. However, there is an extended tail where $\avir \gg 2$, indicating that a significant fraction of the sample are either very short-lived or pressure-confined, and a smaller fraction of sources appear to be unstable to gravitational collapse ($\avir < 1$) unless supported by magnetic fields.

We have adopted a definition of the virial parameter that is appropriate for spherically symmetric sources with a density profile $\rho(r) \propto r^{-2}$, which is the same formulation adopted by the GRS sample, and we remind the reader that this is a very imprecise measure of the state of equilibrium of objects that have complicated shapes in reality. The shape of the distribution of virial ratios for the CHIMPS clumps is of a similar shape to that of the GRS molecular clouds, but with a systematic shift of a factor of $\sim$10 to higher values. If the gas traced by the $^{13}$CO (3--2) is associated with the higher-density regions that are more likely to contain sites of ongoing star formation, then this gas could be expected to carry a relative excess of kinetic energy as a result of feedback processes.

If the CHIMPS clumps are tracing an intermediary stage in the evolution of cells of denser gas from within a molecular cloud to the dense clumps that predominantly go on to form stars, then the distributions of virial ratios would appear to show a surprising movement. It could be that only a relatively small fraction of the CHIMPS clumps are gravitationally collapsing, and are those that comprise the intermediate stage in the transition to the dense star-forming clumps, observable as dust clumps.

Comparison of the sample of CHIMPS clumps that have a one-to-one match with an ATLASGAL clump (the sample of Fig.~\ref{fig:CHIMPSvsATLASGAL}) to the remainder of the whole (reliable) sample, the former do have lower virial parameters than the latter, with $\avir = 1.7^{+1.4}_{-0.7}$ compared with $\avir = 2.5^{+2.3}_{-1.2}$, where the figures give the median value of the distribution and the first and third quartiles. The subsample of matches in which the ATLASGAL clump is associated with a tracer of high-mass star formation has lower values still, with $\avir = 1.4^{+1.3}_{-0.4}$. This is in agreement with \citet{Kauffmann+13}, who found that low virial parameters are often associated with high-mass star formation. However, many of the CHIMPS clumps with the lowest values of $\avir$ are not associated with any ATLASGAL clumps at all, indicating that a low $\avir$ is not a signpost for high-mass star formation in itself.

However, the majority of $^{13}$CO (3--2) clumps from the matched CHIMPS-ATLASGAL sample still have significantly higher virial parameters than those determined by \citet{Urquhart+18} for their dust-traced counterparts, and this might be explained by the larger extent of the clumps (and hence larger radii, leading to larger virial parameters) as traced by the molecular gas. Ground-based observations of thermal dust continuum, such as those of ATLASGAL, must necessarily be spatially filtered in the data reduction process (and ATLASGAL is no exception), and the removal of extended emission might contribute to the discrepancy in virial ratios, in addition to the difference in sensitivities.

The distributions of the crossing timescales and the free-fall timescales of CHIMPS clumps, GRS clouds, and ATLASGAL clumps are compared in Fig.~\ref{fig:timescales}, where:

\begin{equation}
    t_\mathrm{cross} = 2 R_\mathrm{eq}/\sigma_v,
\end{equation}

\noindent and

\begin{equation}
    t_\mathrm{ff} = \sqrt{\frac{3 \pi}{32 G \mu m_\mathrm{p} \bar{n}(\mathrm{H}_2)}}.
\end{equation}

\noindent The peaks of the CHIMPS clump and GRS molecular-cloud crossing time distributions are separated by almost an order of magnitude, suggesting that the GRS clouds are about 3--5 times as long-lived as the CHIMPS clumps which are contained within them. Otherwise, the shapes of the distributions are remarkably similar; they both have a range $\gtrsim 1$ order of magnitude, and standard deviations of $\sim 0.3$ dex. In terms of their free-fall timescales, the CHIMPS clumps and GRS clouds are much more closely matched, owing to their similar densities. The ATLASGAL clumps have much shorter timescales than the CHIMPS clumps and GRS clouds in both cases, suggesting that they evolve much more rapidly. For both measures, the CHIMPS clumps have timescales on the order of a few Myr, that are easily long enough to form high-mass YSOs and compact \ion{H}{ii} regions, which have lifetimes of up to a few $10^5$ years, and the most luminous high-mass YSOs have lifetimes of $\sim 7 \times 10^4$ years \citep{Mottram+11b}.

The difference in crossing timescales of the GRS clouds and the CHIMPS clumps gives a suggestion of the dynamic internal substructure of a cloud. Through the collation of observational evidence, \citet{Elmegreen00} found that star formation in molecular clouds operates over the space of only one or two dynamical timescales. If the structures seen in CHIMPS represent the denser interiors of the large-scale clouds seen in the GRS, then this would appear to suggest that while molecular clouds are transient objects in themselves, their interiors are changing on even shorter timescales. This is consistent with what is expected from the size--linewidth relation, and implies a continuity between the turbulence inherited the largest scales from the ISM, and on the small-scales from gravitational collapse.

\subsection{Trends with Galactocentric radius}

It is well known that a number of ISM properties vary as a function of Galactocentric radius. For example, the metallicity \citep{Caputo+01,Luck+Lambert11}, molecular-to-atomic gas ratio \citep[e.g.,][]{Sofue+Nakanishi16}, and interstellar radiation field \citep[e.g.,][]{Popescu+17}, have all been measured to decrease as a function of increasing Galactocentric radius. The dust temperature within clumps has been found to increase moderately \citep{Urquhart+18} as a function of Galactocentric radius, although \citet{Marsh+17} found that the dust temperature on large scales decreases with radius, and \citet{Roman-Duval+10} found that the mean excitation temperature for molecular clouds also decreases. 

\begin{figure*}
	\includegraphics[width=17cm]{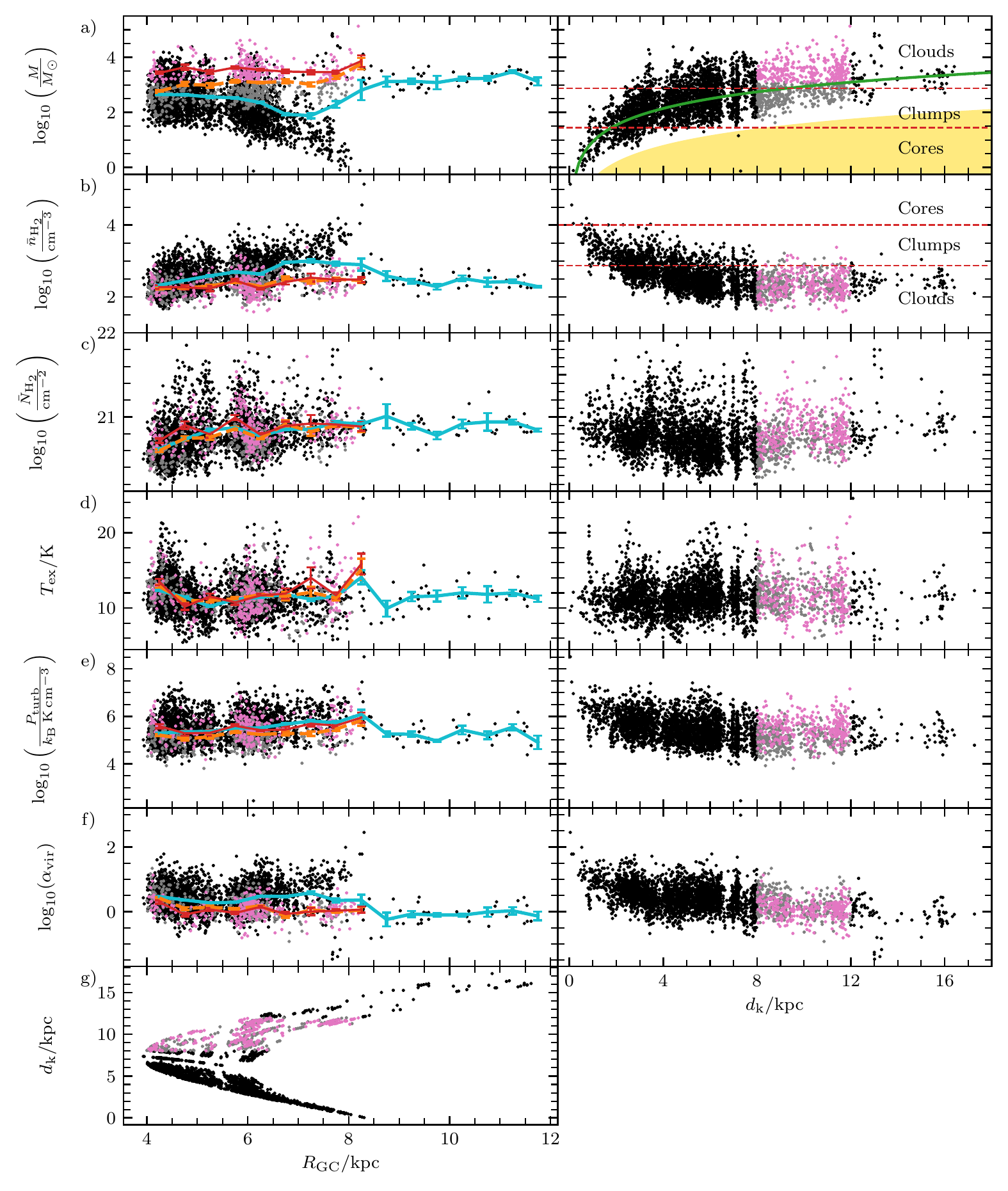}
	\caption{Various properties measured for the CHIMPS clumps as functions of both Galactocentric (left column) and heliocentric distance (right column): a) total clump mass; b) volume-averaged number density c) mean of the voxel column density distribution associated with the clump; d) mean excitation temperature; e) turbulent pressure; f) virial parameter; and g) heliocentric distance. In each case, the distribution of black points shows the full sample (described in Sect.~\ref{sec:properties}), while the grey points show a heliocentric distance-limited subsample, lying between $8 \leq d_\mathrm{k} \leq 12$\,kpc. The pink points show clumps from the distance-limited subsample that lie above the mass completeness limit at 12\,kpc. On the Galactocentric distance distributions, the blue trend lines show the mean values of clumps in 0.5\,kpc-wide bins; while the orange trend lines show the same averages, but for the distance-limited subsample; and the thin red line shows the trend for the mass-complete distance-limited subsample. In the heliocentric-distance column, the dashed red lines denote approximate boundaries between the \citet{Bergin+Tafalla07} `cloud', `clump', and `core' definitions. The green curve in the $M$-$d_\mathrm{k}$ panel denotes the 90\% completeness limit for compact sources, while the yellow shaded region denotes the parameter space below the 5$\sigma$ survey sensitivity.}
	\label{fig:properties_vs_distance}
	\vfill
    \clearpage
\end{figure*}

In Fig.~\ref{fig:properties_vs_distance}, various physical properties of the CHIMPS clumps are shown as a function of both Galactocentric (left column) and heliocentric (right column) distances. We overlay the trends with Galactocentric distance, averaged over 0.5\,kpc-wide bins, in blue and orange, determined from the full and distance-limited samples, respectively, while the thin red trend-line shows the variation in sources from the distance-limited sample that lie above the first-order mass completeness limit at 12\,kpc (from Eq.~\ref{eq:masscompleteness}). To test whether any of these quantities show systematic trends with radius, we performed a linear least-squares fit to the binned quantities from the mass-complete distance-limited sample.

\afterpage{\clearpage}

We see evidence for a shallow increase in the average clump density as a function of Galactocentric radius, with densities increasing by a factor of $\sim2$, over the 4\,kpc range probed. Fig. \ref{fig:properties_vs_distance} would appear to show an increasing trend of excitation temperature with Galactocentric radius, and the linear least-squares fit finds that temperatures increase by $\sim2.5$\,K over the 4\,kpc range. We note that this particular sample of clumps contains a number of sources from with the W49 -- a well known example of an `extreme' star-forming region -- to which the apparent `spike' in $T_\mathrm{ex}$ at $R_\mathrm{GC} \sim 8$\,kpc in Fig.~\ref{fig:properties_vs_distance} can be attributed. This region contains several compact \ion{H}{II} regions, and subsequent heating of the molecular gas might explain this rise in excitation temperature traced by CHIMPS, increasing the apparent strength of the shallow underlying trend with Galactocentric distance. A similar `spike' was seen in the radial excitation-temperature distribution of GRS molecular clouds, which \citet{Roman-Duval+10} attributed to the inclusion of W51 -- another extreme star-forming region -- within the sample. However, the slope in the Galactocentric radius--$T_\mathrm{ex}$ slope does not have a high statistical significance, and the null hypothesis that there is no slope can not be rejected (with a two-sided $p$-value of 0.15). We note that the CHIMPS survey does not cover W51, which is located just beyond the high-longitude end of the survey area.

We do not see any significant systematic changes in the clump masses, mean column densities, mean excitation temperatures, turbulent pressures, or virial parameters over the same range. The apparent slight increase in mean cloud mass with Galactocentric radius in the distance-limited sample disappears once the 12-kpc mass completeness limit is applied, although the distance-limited sample is reduced from 671 to 302 sources in this way. The mean column density and excitation temperature are less dependent on the heliocentric distance, although there is still a dependence, since the former quantity relies upon an optical depth estimation, which is less reliable where C$^{18}$O emission is not detected, and the latter may vary if the beam filling factor varies across the distance range. In all cases, the scatter in the distributions of each property dominates over any global trends.

\subsection{Variations between arm and inter-arm regions} \label{sec:arms}

\begin{figure*}[t]
	\resizebox{\hsize}{!}{\includegraphics{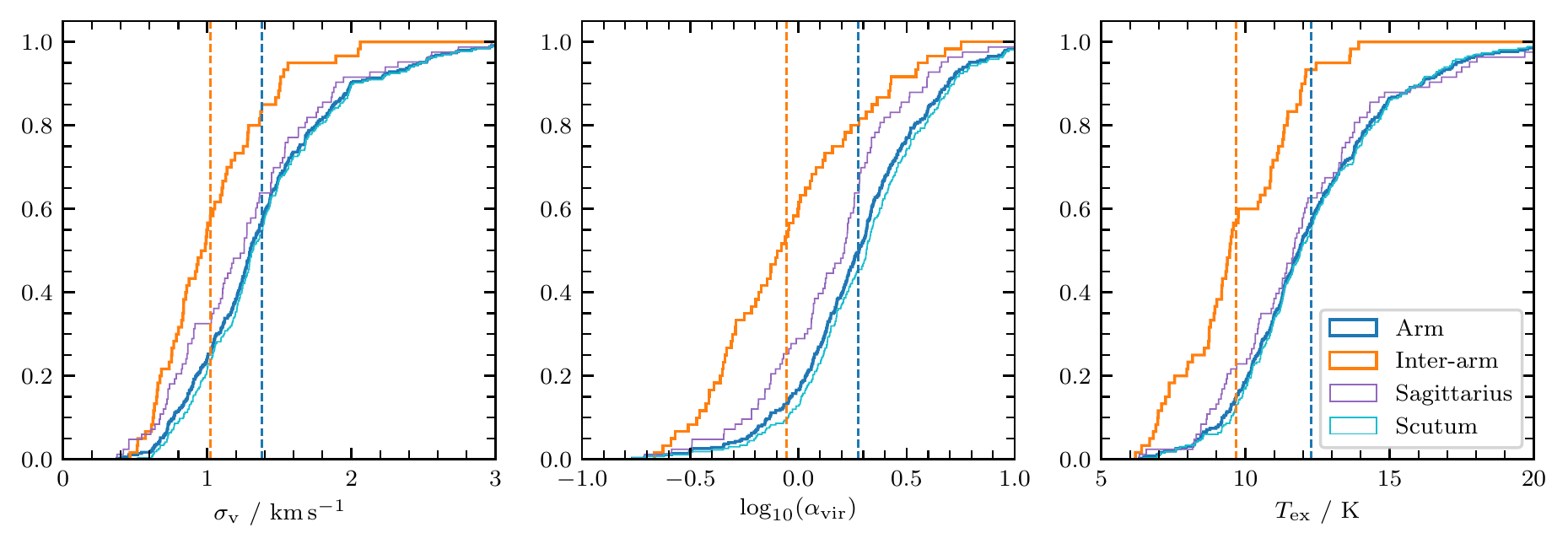}}
    \vspace{-5mm}	
    \caption{Cumulative distributions of the linewidths (left panel), mean excitation temperatures (middle panel), and the logarithm of the virial parameters (right panel) for spiral-arm and inter-arm clumps are shown as the solid blue and orange lines, respectively. The samples are distance-limited with $6\,\mathrm{kpc} \leq d_\mathrm{k} < 9$\,kpc, and exclude clumps with masses below the nominal completeness limit at 9\,kpc. The distributions for the Sagittarius and Scutum arms, which contribute to the spiral-arm distribution, are also shown as the thin purple and cyan lines, respectively. The dashed vertical lines show the means of the spiral-arm and inter-arm distributions.}
    \label{fig:armprops}
\end{figure*}

Only relatively recently have advances in interferometer facilities at millimetre wavelengths enabled molecular-cloud populations to be studied in detail in external galaxies. Using data from the PdBI Arcsecond Whirlpool Survey (PAWS), \citet{Colombo+14} found that giant molecular clouds in the spiral arms of M51 have higher velocity dispersions than in the inter-arm regions, and that the most-massive molecular clouds are found exclusively within the spiral arms. Studies of variations in the efficiency of star formation within our own Galaxy have found only moderate enhancements associated with the spiral arms \citep[e.g.][]{Moore+12,Eden+15}, but that these enhancements can be attributed to individual extreme star-forming regions, which happen to be located within spiral arms. The mass and size distributions of molecular clouds identified within the COHRS survey ($^{12}$CO 3--2) were found to show no significant differences \citep{Colombo+19}, although the authors point out that the tracer may not be sensitive to the most-massive clouds. A greater dynamic range of mass ought to be accessible with the CHIMPS data, owing to the lower opacity tracer.

We examined another distance-limited subsample of 416 CHIMPS clumps, with a range in heliocentric distances of $6\,\mathrm{kpc} \leq d_\mathrm{k} < 9\,\mathrm{kpc}$, and with masses greater than the nominal mass completeness limit at 9\,kpc. This distance range was selected to cover sections of the Scutum-Centaurus and Sagittarius spiral arms, and a significant amount of enclosed inter-arm material, as defined by the \citet{Reid+14} spiral-arm model. We divided this sample into spiral-arm and inter-arm subsets (based on the \citealt{Reid+14} classification), and compared the distributions of physical properties between the samples using a series of two-sample Anderson-Darling (A-D) tests, and by testing whether the mean values of the distributions were significantly different. 

Although we find that the distribution of $\log_{10}(\bar{n}(\mathrm{H}_2) / \mathrm{cm}^{-3})$ is significantly different between the arm and inter-arm regions according to the A-D statistic, with higher mean densities in the inter-arm regions, they do not have significantly different mean values. However, the distributions of $\sigma_{v}$, $\log_{10}(\alpha_\mathrm{vir})$, and $T_\mathrm{ex}$ (shown in Fig.~\ref{fig:armprops}) all show significant differences in both the A-D tests, and with significant differences in their mean values, with lower mean values in each case in the inter-arm sample. The A-D statistic for the arm-interarm comparison of each of these quantities gives a $< 0.1\%$ probability that the null hypothesis that the two samples were drawn from the same distribution is correct. The differences in the mean values of these three quantities are $5.8\,\sigma$, $6.1\,\sigma$ and $8.0\,\sigma$, respectively, where $\sigma$ is the quadrature sum of the standard errors of the two subsamples.

The difference in linewidth between spiral arm and inter-arm regions is comparable to the findings of larger linewidths for molecular clouds located within M51's spiral arms \citep{Colombo+14}, and to the molecular clouds identified within the smoothed particle-hydrodynamics simulation of \citet{Duarte-Cabral+Dobbs16}. The latter study attributes the higher velocity dispersions of clouds within the spiral arms to the more chaotic nature of those environments, with an increased rate of cloud collisions and interactions \citep{Dobbs+15}. In such an environment, the emission associated with a particular molecular cloud is more likely to overlap in position--position--velocity space, and therefore introduces an apparent increase in velocity dispersion. The lower virial parameters in the inter-arm regions is reminiscent of the simulations of tidally-induced spiral arms of \citet{Pettitt+18}, who found that the least-bound GMCs show a strong preference for residing within spiral arms, while the most bound ones exhibit a much weaker correspondence. Some of the bound inter-arm GMCs were found to be remnants of larger complexes within the spiral arms, although there were also a population that had formed in situ. In this case, the tidal spiral arms serve to protect the GMCs from the shear induced by the differential rotation of the disc, which quickly destroys unbound clouds in the inter-arm regions.

Further tests were performed by breaking the spiral-arm sample into its constituent populations attributed to the Sagittarius and Scutum-Centaurus spiral arms, which are made up of 83 and 266 clumps, respectively. We find that the distribution of $\log_{10}(R_\mathrm{eq})$ is significantly (at the 3.0\,$\sigma$ level) larger in the Sagittarius arm than in the Scutum-Centaurus arm, and the $p$-value from the A-D test indicates a $< 0.1\%$ probability that the two samples are drawn from the same underlying distribution. Clumps in the Sagittarius arm are, on average, 20\% larger than their counterparts in the Scutum-Centaurus arm. The application of the mass-completeness limit (the origin of which is described in Appendix~\ref{app:completeness}) does not account for the preferential sizes isolated by {\sc FellWalker}, and so might play some role in artificially altering the apparent clump sizes as a function of distance. However, the median distance to sources in the Sagittarius arm sample is 8.0\,kpc compared with 7.1\,kpc in the Scutum-Centaurus arm, and so this finding is unlikely to be entirely explained by being able to detect more diffuse material at larger radii in the nearer sample. Source crowding within the Scutum-Centaurus arm might mean that {\sc FellWalker} is also playing a role here, though it is difficult to conceive of how the higher levels of source-crowding would decrease the source size.  Further, the algorithm contains a `cleaning' automaton\footnote{For a description of how {\sc FellWalker}'s cellular automata merge adjacent clumps (as controlled by the CleanIter parameter), see http://www.starlink.ac.uk/docs/sun255.htx/sun255se2.html} that inclines to merge adjacent clumps, which would tend to increase the source size in crowded regions.

The mean value of $\log_{10}(\alpha_\mathrm{vir})$ for clumps in the Sagittarius spiral arm is also significantly lower (at the $3.3\,\sigma$ level) than than that of Scutum-Centaurus arm, and the underlying distributions are also significantly different according to the A-D test. The lower virial parameters in the Sagittarius spiral arm present a particularly interesting case because the Sagittarius arm has been found to contain molecular clouds with an enhanced excitation temperature relative to its surroundings in the GRS survey \citep{Roman-Duval+10}, as well as containing a high density of high-mass star-forming regions \citep{Urquhart+14}, and exhibiting high values of the ratio of IR luminosity to clump mass \citep{Moore+12,Eden+15}, implying a higher star-formation efficiency. Such low virial parameters in the Sagittarius arm might be expected if low virial parameters are indeed characteristic of high-mass star-forming regions, as suggested by \citet{Kauffmann+13}. However, it is unclear whether or not the blending of sources within the very crowded W43 star-forming region -- present within this Scutum-Centaurus sample -- could produce an acute artificial increase in the virial parameters measured there.

\section{Summary and conclusions} \label{sec:conclusion}

By using an LTE analysis to combine the $^{13}$CO (3--2) and C$^{18}$O (3--2) CHIMPS data with COHRS $^{12}$CO (3--2) data, we have determined the excitation temperatures, optical depths, and column densities of $^{13}$CO (3--2) emission throughout the CHIMPS survey volume on a voxel-by-voxel basis. We have performed a source extraction on the $^{13}$CO (3--2) data using the {\sc FellWalker} algorithm, identifying a robust sample of 3664 molecular clumps out of a total of 4999 candidates, and within the literature we have determined kinematic distances to those clumps through associations with dense clumps and molecular clouds. 

We have examined the physical properties of the 3553 of the population of molecular clumps for which we have a distance determination and consistent coverage with CHIMPS and COHRS data, and our main findings are as follows.

\begin{enumerate}[i)]

\item The masses, radii, densities, excitation temperatures. and dynamical lifetimes of the CHIMPS clumps fall in an intermediate parameter space between molecular clouds (as traced by $^{13}$CO $J$=1--0 in the Galactic Ring Survey) and dense clumps (traced by thermal dust continuum in ATLASGAL). We interpret this as evidence that the CHIMPS clumps represent an intermediate phase of molecular cloud substructure between molecular clouds and dense clumps (traced by thermal-dust emission from ATLASGAL) in terms of the derived volume-averaged densities, and the in mass-radius plane.

\item The derived mean clump excitation temperatures range between approximately 5--25 K throughout the surveyed volume, with a mean value of 11.5 K.

\item The median turbulent pressure is a factor of $\sim 60$ larger than the thermal pressure.

 \item The size--linewidth relation of the full sample of CHIMPS clumps is $\sigma_v \propto R_\sigma^{0.41}$, similar to the canonical relationship for Milky Way molecular clouds, but we find a steeper slope of $\sigma_v \propto R_\sigma^{0.63}$ when considering a distance-limited sample that ought to be more robust against distance-related biases. The size--density relationship is shallower than typically observed, with $\sigma_v \propto \bar{n}(\mathrm{H}_2)^{-0.84}$ and $\sigma_v \propto \bar{n}(\mathrm{H}_2)^{-0.68}$ for the full and distance-limited samples, respectively.
 
 \item We do not see any evidence for a significant systematic trend with Galactocentric distance of the clump masses, radii, mean column densities, excitation temperatures, turbulent pressures, or virial parameters. There is a shallow trend for the average volume density to increase as a function of Galactocentric distance, with a factor of $\sim 2$ increase over the probed range ($4 < R_\mathrm{GC} < 8$\,kpc).
 
 \item Comparison of the physical properties of clumps located within spiral arms and inter-arm regions reveals that clumps residing within inter-arm regions have, on average, lower velocity dispersions, virial parameters, and excitation temperatures. This difference in linewidths is in agreement with the smoothed particle hydrodynamics simulation of \citet{Duarte-Cabral+Dobbs16}, and the difference in virial parameters is concordant with the findings of \citet{Pettitt+18}.
 
 \item The median value of the radius distribution for clumps within the Sagittarius spiral arm is significantly higher than that of the Scutum-Centaurus spiral arm, though we can not exclude the possibility that this is partially a result of detecting more extended structure in the nearer Sagittarius arm.
 
 \item Even considering the modest variations in some properties found that vary with Galactocentric distance, or between spiral-arm and inter-arm regions, the variation in all properties from clump-to-clump is far greater than any systematic environmental dependence. This suggests that the impact of Galactic environment across the inner disc upon molecular-clump properties is rather minimal.

\end{enumerate}

We have made the $^{13}$CO (3--2) source catalogue (Table \ref{tab:FWcat}), along with a catalogue listing all of the derived physical properties (\ref{tab:properties}) publicly available at \url{https://doi.org/10.11570/19.0028}. We have also made available the source masks, along with the corresponding $^{13}$CO and C$^{18}$O (3--2) `Region' cubes, $^{13}$CO column density, excitation temperature, and optical depth cubes. Further details about these can be found in Appendix~\ref{app:dataproducts}.

In future work we will exploit synthetic observations produced from molecular-cloud simulations, such as those of \citet{Penaloza+17}, to refine our determination of the properties of the molecular gas, incorporating non-LTE methods. We will also expand our study of clump-property variations over a much wider range of Galactic environments as part of the CHIMPS2 Large Program (Eden et al. in prep).

\begin{acknowledgements}
The authors thank the anonymous referee for their constructive comments that have improved the quality of the paper. The authors also thank Ana Duarte-Cabral, Paul Clark and Camilo Pe\~{n}aloza for many helpful discussions relating to this work. AJR would like to thank the STFC for postdoctoral support under the consolidated grant number ST/N000706/1, as well as funding for a PhD studentship. SER acknowledges support from the European Union's Horizon 2020 research and innovation programme under the Marie Sk{\l}odowska-Curie grant agreement \# 706390. This research made use of Astropy\footnote{\url{https://astropy.org/}}, a community-developed core Python package for Astronomy \citep{astropy2013, astropy2018}, as well as iPython\footnote{\url{https://ipython.org/}}, an enhanced interactive Python \citep{Perez+07}, and the Python packages NumPy\footnote{\url{https://numpy.org/}} \citep{Oliphant06}, SciPy\footnote{\url{https://scipy.org/}} \citep{Jones+01}, and Matplotlib\footnote{\url{https://matplotlib.org/}} \citep{Hunter07}. This publication has also made use of TOPCAT\footnote{\url{http://www.star.bris.ac.uk/~mbt/topcat/}}, an interactive graphical viewer and editor for tabular data software \citep{Taylor05}, and SAOImageDS9\footnote{\url{http://ds9.si.edu/}} \citep{Joye+03}, an astronomical imaging and data-visualization application. The James Clerk Maxwell Telescope has historically been operated by the Joint Astronomy Centre on behalf of the Science and Technology Facilities Council of the United Kingdom, the National Research Council of Canada and the Netherlands Organisation for Scientific Research. This research used the facilities of the Canadian Astronomy Data Centre operated by the National Research Council of Canada with the support of the Canadian Space Agency, and has made use of the NASA Astrophysical Data System. The Starlink software \citep{Currie+14} is currently supported by the East Asian Observatory.
\end{acknowledgements}


\bibliographystyle{aa}
\bibliography{References_CHIMPSII}

\begin{appendix}

\section{Data products} \label{app:dataproducts}

We present a number of public data products as a result of this study, which can be found on the CANFAR data archive\footnote{\url{https://doi.org/10.11570/19.0028}}, and which are detailed below.

\begin{itemize}
    \item We present two catalogues: i) the {\sc FellWalker} source catalogue, detailing the observed properties of discrete sources identified from in the $^{13}$CO (3--2) data, including angular sizes, linewidths, and integrated intensities calculated at the native (15 arcsecond) resolution; ii) the catalogue of physical properties derived from the LTE analysis of Sect.~\ref{sec:LTE} at an angular resolution of 27.4 arcseconds, including masses, excitation temperatures, virial parameters, and reliability flags.
    
    \item We make the {\sc FellWalker} source masks available as a series of ten position-position-velocity cubes, each given by a `Region' number, and the pixel value of each source corresponds to an identifier in the two catalogues mentioned above.
    
    \item We also make available the $^{13}$CO and C$^{18}$O (3--2) intensity, $N_{13}$(total), $\tau_{13}$, and $T_\mathrm{ex}$ cubes, which correspond to each of the ten `Region' cubes.
    
    \item For each source, we also present a cutout image in PDF format, showing the integrated $^{13}$CO (3--2) intensity (on the corrected antenna temperature, $T_\mathrm{A}^{*}$ scale), and outlining the source in its surroundings.
    
\end{itemize}

\section{Completeness tests} \label{app:completeness}

To estimate the completeness of the source sample extracted from the CHIMPS data, a number of tests were carried out in which fake sources were injected into a sample of the $^{13}$CO (3--2) data, and extracted in the usual manner. A 0.7 square-degree mosaicked cube of six individual $^{13}$CO (3--2) cubes, centred on $\ell = 30.5\degr, b=0.0\degr$ was chosen from the publicly 
available data (see \citealt{Rigby+16}) as a representative subsample of the survey, with a mean rms of 0.50\,K, and standard deviation of 0.11\,K. This cube also covers the most crowded line of sight within the survey, which includes the W43 star-forming region, and so ought to present the most difficult source-extraction conditions.

Four sets of source injection experiments were carried out, in which three-dimensional Gaussian-profiles were injected at integer peak intensities ranging from $T_\mathrm{A}^{*} = 1$ to 25\,K into the test cube. A total of 10,000 sources were injected into the cube at each peak $T_\mathrm{A}^{*}$ value, made up of 25 realisations of 400 randomly positioned sources, in order to avoid source crowding, which makes re-identification of the injected sources difficult and introduces non-linearity. The four experiments involved the injection of sources of different sizes, with FWHM extents in the $\ell, b, v$ axes of $3 \times 3 \times 2$, $3 \times 3 \times 4$, $6 \times 6 \times 3$ and $10 \times 10 \times 4$ pixels before smoothing, respectively, approximately covering the parameter space of the deconvolved sizes of the extracted sources. 

\begin{figure}
	\resizebox{\hsize}{!}{\includegraphics{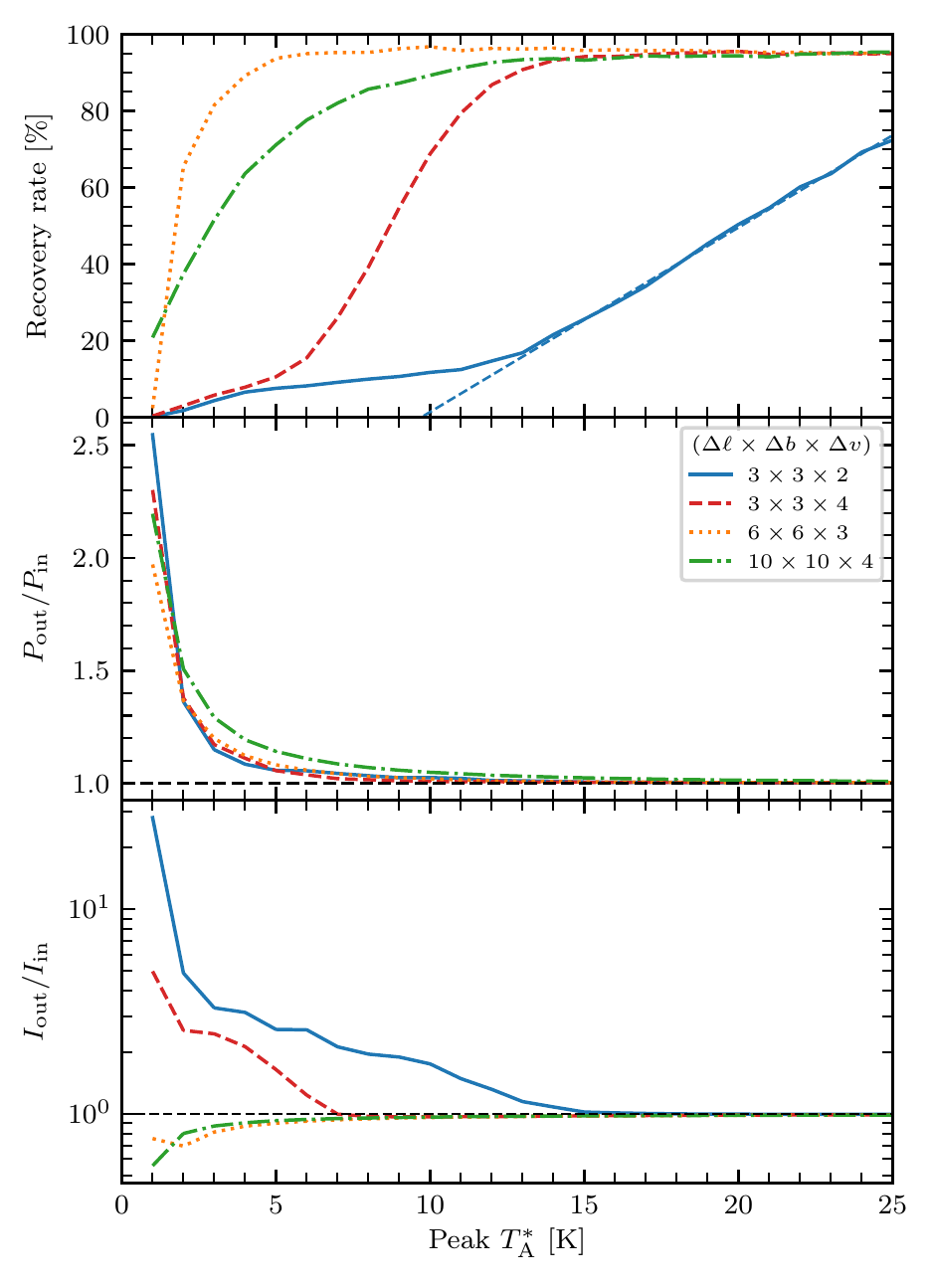}}
    \vspace{-5mm}	
	\caption{Results of the completeness tests, comparing the recovered quantities for 10,000 randomly injected three-dimensional-Gaussian sources with sizes typical of the CHIMPS $^{13}$CO (3--2) sources as a function of the injected peak $T_\mathrm{A}^{*}$ values. The panels are: fraction of sources recovered (top panel); the median ratio of the recovered to injected peak brightness temperature (middle panel); and the median ratio of the recovered-to-input integrated intensity (bottom panel).}
	\label{fig:completeness}
\end{figure}

The results of the completeness tests are presented in Fig.~\ref{fig:completeness}. In the top panel, the recovery rate as a function of the injected peak $T_\mathrm{A}^{*}$ is shown. The recovery rate is defined as the fraction of the 10,000 sources per peak $T_\mathrm{A}^{*}$ that lie within the output catalogue, that are identified with a peak position lying within half of the effective-resolution element in $\ell, b$, and within one pixel in $v$ of the injected position. The most-compact sources are not recovered well until they are approaching the maximum input value of $T_\mathrm{A}^{*} = 25$\,K, which is approximately the brightest emission seen in the full survey. The intermediate-sized sources are well recovered, reaching a maximum recovery rate of $\approx 95\%$ by a peak $T_\mathrm{A}^{*} = 5$\,K, corresponding to a peak S/N of 10, while the largest injected sources converge to a 95\% completeness rate with sources with a peak of $T_\mathrm{A}^{*} = 12$\,K. The largest sources are the most likely to coincide with existing emission upon injection, and are therefore the most likely to be subsumed into pre-existing emission features, especially since {\sc FellWalker} contains a `cleaning' algorithm to join up sources with overlapping boundaries.

For all sources, the median ratio of the recovered-to-injected peak $T_\mathrm{A}^{*}$, shown in the middle panel of Fig.~\ref{fig:completeness}, is largest at the lowest injected-peak intensity level. This is a consequence of the low S/N, and {\sc FellWalker} reports the intensity maximum as the peak value within the identified clump volume, preferentially selecting the extreme of the signal-plus-noise distribution (an effect sometimes known as `flux boosting'). The bottom panel shows the median ratio of the recovered-to-injected volume-integrated intensity per source, which also shows an overestimate for the most-compact objects at the lowest intensities. In this case, the recovery of the injected sources is almost impossible, with a recovery of only 14 of the 10,000 injected sources recovered at a peak $T_\mathrm{A}^{*} = 1$\,K, and these are chance alignments in where the injected sources are placed within existing -- and brighter -- emission features that have been identified. Re-identifying the injected sources is much easier for the larger sources, for which the recovered integrated intensity is lower than the injected integrated intensity at low S/N. This is a threshold effect, in which a larger fraction of the intensity at low S/N is considered to be below the noise limit (see Fig. 9 of \citealt{Berry15}). In all cases, the ratios of both the recovered peak $T_\mathrm{A}^{*}$ and integrated intensities to the injected quantities tend to unity as the sources become brighter, and integrated intensities are well reported above the average $\mathrm{S/N} = 5$ detection level for the test cube, which is around a peak $T_\mathrm{A}^{*}=2.5$\,K.

\begin{table}
	\centering
    \caption{Approximate 90\% completeness level, in terms of the peak $T_\mathrm{A}^{*}$ for Gaussian sources of varying FWHM extent in the $\ell,b,v$ axes. For a given source size, the completeness limit varies as a function of distance $M_\mathrm{complete}^{90\%}(d) = \Sigma_0 \, d^{2}$.}
    \begin{tabular}{ccc}
    \hline
    \noalign{\smallskip}
    FWHM size & $T_\mathrm{A}^{*}(90\%)$ & $\Sigma_0$\\
    (pixels) & (K) & ($M_\sun$\,kpc$^{-2}$) \\
    \noalign{\smallskip}
    \hline
    \noalign{\smallskip}
    $3 \times 3\times 2$	& 28 K & 8.9 \\
	$3 \times 3\times 4$	& 13 K & 8.1 \\ 
	$6 \times 6\times 3$	& 5 K  & 8.9 \\ 
	$10 \times 10\times 4 $ & 11 K & 76.0 \\
	\noalign{\smallskip}
    \hline
    \end{tabular}
    \label{tab:completeness}
\end{table}

To estimate a functional completeness limit, the peak $T_\mathrm{A}^{*}$ at which the recovery rate curves reach the 90\% level for each source size was recorded, and converted into a mass by integrating the intensity over the corresponding source profile. In the case of the most-compact sources, the 90\% completeness level is not reached within the tested peak $T_\mathrm{A}^{*}$ range, and so a value of 28 K was recovered by extrapolating the bright end of the recovery-rate curve, shown as the blue dashed line in the top panel of Fig.~\ref{fig:completeness}. This mass, calculated assuming a mean optical depth of 0.36 and excitation temperature of 11.5\,K (the mean values, as determined in Sect.~\ref{sec:LTE}, for clumps with the highest reliability flag), allows the calculation of a 90\% completeness level as a function of heliocentric distance, $M_\mathrm{complete}^{90\%} = \Sigma_0 \, d^{2}$, where $\Sigma_0$ is a normalisation factor and $d$ is the distance in kpc. The $\Sigma_0$ values for the different source sizes are given in Table~\ref{tab:completeness}. The 90\% completeness limit is similar for the compact sources, corresponding to $\sim900\,M_\sun$ at a distance of 10\,kpc, but large sources must be very bright in order to be recovered as a single object.

A caveat in this analysis is that these injected sources do not look like all of the sources in the survey. While there are many structures that are compact, like the just-resolved, circular-profiled Gaussian sources injected, there are also many sources that have complex and irregular shapes, accompanied by irregular intensity profiles. These kinds of sources are extremely hard to replicate, so the best-matching completeness limit should be selected from the most appropriate source type in Table~\ref{tab:completeness} for further analyses. To first order, we recommend that a mean value from the three sets of more-compact sources be chosen to a simple completeness limit, with

\begin{equation} \label{eq:masscompleteness}
M_\mathrm{complete}^{90\%} = 8.7 \ \left(\frac{d}{\mathrm{kpc}}\right)^2 M_\odot,
\end{equation}

\noindent noting the caveats that this completeness limit is only strictly applicable for unresolved sources, and is a conservative figure due to testing in W43 -- the most crowded region within the survey.

\section{Spiral-arm and inter-arm sample statistics} \label{app:armstats}

In Sect.~\ref{sec:arms} we compared the values of several quantities across a number of spiral-arm and inter-arm subsamples, picked from within a mass-complete and distance-limited sample of clumps. Some of the basic statistics for these samples are presented in Table~\ref{tab:arms}. We present the median and mean values for each quantity, along with the standard error upon the mean ($\sigma / \sqrt{N-1}$), alongside $p$-values resulting from a two-sample Anderson-Darling test carried out for the pairs of subsamples that were compared.

\begin{table*}[b]
\centering
\caption{Statistics from the comparison of the spiral-arm and inter-arm subsamples in Sect.~\ref{sec:arms}.}
\begin{tabular}{cccccc}
\hline \hline
\noalign{\smallskip}
Quantity & Subsample & Median & Mean & Standard & A-D \\
 &  & &  & Error & $p$-value \\
\noalign{\smallskip}
\hline
\noalign{\smallskip}
\multirow{4}*{$\log_{10}(M / M_\odot)$} & Arm & 3.236 & 3.292 & 0.018 & \multirow{2}*{9.6\%} \\
 & Int & 3.293 & 3.371 & 0.059 &  \\\cline{6-6}
 & Scu & 3.214 & 3.265 & 0.020 & \multirow{2}*{1.2\%} \\
 & Sag & 3.332 & 3.377 & 0.043 &  \\
\noalign{\smallskip}
\hline
\noalign{\smallskip}
\multirow{4}*{$\log_{10}(R_\mathrm{eq} / \mathrm{pc})$} & Arm & 0.212 & 0.206 & 0.010 & \multirow{2}*{$> 25.0$\%} \\
 & Int & 0.236 & 0.202 & 0.028 &  \\\cline{6-6}
 & Scu & 0.197 & 0.189 & 0.011 & \multirow{2}*{0.1\%} \\
 & Sag & 0.270 & 0.262 & 0.022 &  \\
\noalign{\smallskip}
\hline
\noalign{\smallskip}
\multirow{4}*{$\log_{10}(n(\mathrm{H}_2) / \mathrm{cm}^{-2})$} & Arm & 2.309 & 2.346 & 0.019 & \multirow{2}*{0.1\%} \\
 & Int & 2.437 & 2.510 & 0.057 &  \\\cline{6-6}
 & Scu & 2.300 & 2.353 & 0.023 & \multirow{2}*{24.3\%} \\
 & Sag & 2.310 & 2.327 & 0.034 &  \\
\noalign{\smallskip}
\hline
\noalign{\smallskip}
\multirow{4}*{$\sigma_\mathrm{v}$ / km\,s$^{-1}$} & Arm & 1.297 & 1.379 & 0.030 & \multirow{2}*{$< 0.1$\%} \\
 & Int & 0.973 & 1.024 & 0.054 &  \\\cline{6-6}
 & Scu & 1.316 & 1.411 & 0.034 & \multirow{2}*{1.8\%} \\
 & Sag & 1.256 & 1.277 & 0.067 &  \\
\noalign{\smallskip}
\hline
\noalign{\smallskip}
\multirow{4}*{$\log_{10}(\alpha_\mathrm{vir})$} & Arm & 0.281 & 0.277 & 0.019 & \multirow{2}*{$< 0.1$\%} \\
 & Int & -0.089 & -0.056 & 0.051 &  \\\cline{6-6}
 & Scu & 0.315 & 0.313 & 0.021 & \multirow{2}*{$<0.1$\%} \\
 & Sag & 0.214 & 0.163 & 0.040 &  \\
\noalign{\smallskip}
\hline
\noalign{\smallskip}
\multirow{4}*{$\log_{10}(P_\mathrm{turb} / k_\mathrm{B} \, \mathrm{K\,cm}^{-3})$} & Arm & 5.561 & 5.559 & 0.028 & \multirow{2}*{$>25.0$\%} \\
 & Int & 5.539 & 5.473 & 0.078 &  \\\cline{6-6}
 & Scu & 5.592 & 5.591 & 0.032 & \multirow{2}*{5.1\%} \\
 & Sag & 5.528 & 5.455 & 0.059 &  \\
\noalign{\smallskip}
\hline
\noalign{\smallskip}
\multirow{4}*{$T_\mathrm{ex}$ / K} & Arm & 11.858 & 12.273 & 0.154 & \multirow{2}*{$< 0.1$\%} \\
 & Int & 9.461 & 9.673 & 0.284 &  \\\cline{6-6}
 & Scu & 11.920 & 12.346 & 0.173 & \multirow{2}*{22.5\%} \\
 & Sag & 11.659 & 12.040 & 0.365 &  \\
\noalign{\smallskip}
\hline
\noalign{\medskip}
\multicolumn{5}{p{0.6\textwidth}}{\textbf{Notes.} The subsample names are listed in abbreviated form, and are designated as the spiral arm (`Arm'), inter-arm (`Int'), Scutum-Centaurus arm (`Scu') and Sagittarius arm (`Sag'). We list the two-sample Anderson-Darling test $p$-values.}
\end{tabular}
\vspace{4cm}
\label{tab:arms}
\end{table*}

\end{appendix}

\end{document}